\documentclass{article}

\usepackage{graphicx} 
\usepackage{acronym}
\usepackage{makecell}
\usepackage{booktabs}
\usepackage{amsmath}

\acrodef{ABM}{Agent-Based Modeling}
\acrodef{EBM}{Equation-Based Modeling}
\acrodef{SD}{System Dynamics}
\acrodef{DES}{Discrete Events Simulation}
\acrodef{CA}{Cellular Automaton}
\acrodefplural{CA}{Cellular Automata}
\acrodef{DoS}{Denial of Service}
\acrodef{DDoS}{Denial of Service}
\acrodef{MitM}{Man-in-the-Middle}
\acrodef{SIFF}{Stateless Internet Flow Filter}
\acrodef{AITF}{Active Internet Traffic Filtering}
\acrodef{TVA}{Traffic Validation Architecture}
\acrodef{SCADA}{Supervisory Control and Data Acquisition}
\acrodef{HIL}{Hardware-in-the-loop}
\acrodef{COA}{Cascading Outage Analysis}
\acrodef{IDS}{Intrusion Detection System}
\acrodef{XSS}{Cross-Site Scripting}
\acrodef{P2P}{Peer-to-peer}
\acrodef{CPS}{Cyber-Phisical System}
\acrodef{CPPS}{Cyber-Phisical Power System}
\acrodef{APT}{Advanced Persistent Threat}
\acrodef{UAV}{Unmanned aerial vehicle}
\acrodef{MPC}{Model Predictive Control}
\acrodef{WAMS}{Wide-Area Measurement System}
\acrodef{V2X}{Vehicles to Everything}
\acrodef{PMU}{Phasor Measurement Unit}
\acrodef{IoT}{Internet of Things}
\acrodef{MITM}{Man in the Middle}
\acrodef{DLT}{Distributed Ledger Technology}
\acrodef{UAV}{Unmanned Aerial Vehicle}
\acrodef{WSN}{Wireless Sensor Network}
\acrodef{GAMES}{General Agent Model for the Evaluation of Security}
\acrodef{PNPSC}{Petri Nets with Players, Strategies, and Costs}
\acrodef{SDN}{Software-Defined Networking}
\acrodef{WAF}{Web Application Firewall}
\acrodef{IPS}{Intrusion Prevention System}
\acrodef{IDS}{Intrusion Detection System}
\acrodef{EDR}{Endpoint Detection and Response}

\newcommand{\defaultscale}{0.8}

\title{Simulation in Cybersecurity: Understanding Techniques, Applications, and Goals\\ (Working Paper)}

\author{Luca Serena, Gabriele D'Angelo,\\ Stefano Ferretti, Moreno Marzolla\\University of Bologna\\Bologna, Italy\\
\texttt{\{luca.serena2,g.dangelo,s.ferretti,moreno.marzolla\}@unibo.it}}
\date{}

\begin{document}
\maketitle

\begin{abstract}
Modeling and simulation are widely used in cybersecurity research to assess cyber threats, evaluate defense mechanisms, and analyze vulnerabilities. However, the diversity of application areas, the variety of cyberattacks scenarios, and the differing objectives of these simulations makes it difficult to identify methodological trends. Existing reviews often focus on specific modeling techniques or application domains, making it challenging to analyze the field as a whole.

To address these limitations, we present a comprehensive review of the current state of the art, classifying the selected papers based on four dimensions: the application domain, the types of cyber threats represented, the simulation techniques employed, and the primary goals of the simulation. The review discusses the strengths and limitations of different approaches, identifies which cyber threats are the most suited for simulation-based investigations, and analyzes which modeling paradigms are most appropriate for specific cybersecurity challenges.
\end{abstract}


\section{Introduction}

The shift towards digitalization and the increasing interconnectedness of modern technological devices had a remarkable impact on society.
On one hand, several tasks have been automated and made more efficient; on the other, many systems are now exposed to cyberattacks, which have become a significant threat across various sectors~\cite{burger2019estimating}. Attackers may pursue different goals, such as provoking operational disruption, inflicting financial harm, and stealing confidential information. To achieve this, they exploit vulnerabilities in the target systems, which are any type of weakness or flaw in a system's design, implementation, configuration, or usage.


Among the various categories of cyberattacks, some stand out for their frequency and potential for disruption.
One of the major cybersecurity concerns is \ac{DoS}~\cite{cetinkaya2019overview}, where the targeted system is overwhelmed with requests at a rate it can not sustain, making service unavailable to legitimate users. The impact of~\ac{DoS} can be further increased if multiple compromised devices participate in the attack -- a scenario commonly referred to as a \ac{DDoS} -- making it harder to identify and isolate the responsible parties.
Other attacks aim at gaining unauthorized access or to manipulate the operations of a target system, typically by making the victim execute malicious code through either vulnerable web interfaces or malware deployment. 
Since communication channels are the primary medium through which cyberattacks spread, it is essential to ensure confidentiality, integrity, and authenticity of data transmissions. If channels are not properly secured through cryptographic techniques, these properties can be undermined by a \ac{MitM}~\cite{conti2016survey}, who can intercept, alter, or relay communication between two parties without their knowledge. This allows for eavesdropping on sensitive information, modifying transmitted data, or injecting malicious commands.


The increasing adoption of technologies such as the~\ac{IoT}, \ac{DLT}, and cloud services further expands the attack surface.
In transportation systems, cyber vulnerabilities can compromise the safety and functionality of autonomous vehicles and traffic control networks. In energy and utility infrastructures, attacks on supervisory control and data acquisition systems can cause power outages or water supply failures, affecting millions of people.
Thus, cyberattacks can lead to severe economic, operational, and social consequences, requiring appropriate countermeasures to mitigate risks and damage.
Given the significant impact of cyber threats, it is important to develop methods and tools to assess vulnerabilities, measure the effectiveness of defenses, and evaluate resilience to attacks.
Modeling and simulation is a powerful tool to analyze complex threat scenarios, test mitigation strategies in controlled environments, and gain deeper insights into system behavior under adversarial conditions.

This paper examines the role of simulation in cybersecurity by reviewing and categorizing the research activities of the last~$25$ years, to address the following research questions (RQ):

\begin{description}
\item[RQ1:] \textbf{Which application domains are the most frequently considered by cybersecurity simulations?} 
Since cybersecurity is relevant for all digital applications, this questions aims to identify which sectors are the most represented in the scientific literature.

\item[RQ2:] \textbf{Which types of cyber threats are mostinvestigated using simulation?} 
Different cyber threats are characterized by distinct challenges in terms of modeling and analysis. This questions examines which attack vectors are investigated through simulation and the reasons behind the research approach.

\item[RQ3:] \textbf{Which modeling and simulation techniques are employed in cybersecurity research?} 
Since many different modeling paradigms exist, this question aims at understanding which ones are most suitable for investigating each type of threat or application domain, considering aspects such as computational constraints, the required level of detail, and the need to integrate stochasticity or dynamic behaviors.

\item[RQ4:] \textbf{What are the primary goals of modeling and simulation in cybersecurity research?} 
Most papers in this area share a common objective, such as evaluating the impact of an attack under specific conditions, but there are important differences in the focus of the simulation. Some studies concentrate on defensive measures, others on identifying critical threshold values, analyzing economic consequences, or examining the propagation of malware. This question aims to classify the goals pursued by researchers, and understand how they influence the simulation design choices.

\end{description}

\begin{figure}[ht]
\centering\includegraphics[scale=\defaultscale]{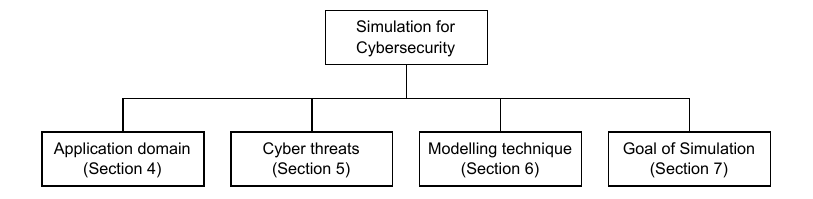}
\caption{Taxonomy of the review.}
\label{taxonomy}
\end{figure}

To this aim, we classify the research papers across four dimensions (Figure~\ref{taxonomy}):

\begin{itemize}
    \item \textit{Application domain}. Cybersecurity risks vary across different sectors, each with specific technological features and potential attack vectors. Simulation enables domain-specific vulnerability analysis, allowing security strategies to be tailored to the operational and structural characteristics of the investigated system.
    \item \textit{Cyber Threat considered}. The term ``cybersecurity'' encompasses a wide range of threats, which differ in the vulnerabilities being exploited, target and objective of the attack, available means, origin (i.e., internal or external), and complexity.  
    \item \textit{Modeling technique}. Several M\&S paradigms are available, depending on the required level of granularity, availability of computing resources, and the complexity of the scenario under study.
    \item \textit{Goal of the investigation}. The goal of simulation-based cybersecurity studies can vary depending on several factors. In most cases, potential weaknesses are already known, and the goal is to examine the level of resilience of the system under adversarial conditions, the effectiveness of certain mitigation strategies, or the level of impairment that is caused by attacks. Conversely, simulations might help to identify vulnerable endpoints where cyber-induced failures would cause the most damage, or generate syntetic workloads to train~AI models to detect malicious patterns that could be symptoms of attacks.
\end{itemize}

This paper is organized as follows. In Section~\ref{sec:related} we present some review papers that discussed simulation papers that have discussed M\&S in the field of cybersecurity.
In Section~\ref{sec:methodology} we define the methodology that has been adopted to collect and categorize the references; we analyze the data from the quantitative point of view to highlight research trends that help to understand how this research area is evolving. 
Sections~\ref{sec:domain} through~\ref{sec:goals} are devoted to the analysis of the scientific literature that has been discussed along the four main dimensions. Finally, conclusions and general remarks will be provided in Section~\ref{sec:conclusions}.

\section{Related Works}\label{sec:related}

\begin{table}[ht]
\centering
\caption{Comparison of this review and others from the literature.}
\small
\begin{tabular}{|l|c|p{0.4\textwidth}|c|}
\hline
\textbf{Study} & \textbf{Year} & \textbf{Survey Focus} & \textbf{N. of Papers} \\
\hline
\cite{yohanandhan2020cyber} & 2020 & Cyber-physical power systems & 67 \\
\cite{vestad2024survey} & 2024 & Agent-based modeling & 39 \\
\cite{jawad2021modeling} & 2021 & Cyber-physical systems, ICS & 31 \\
\cite{chowdhury2021cyber} & 2021 & Critical infrastructures & 68 \\
\cite{kavak2021simulation} & 2021 & Training, risk analysis, testbeds, human behavior & 50 \\
\cite{engstrom2022two} & 2022 & Cause-effect cyberattack simulations & 11 \\
\cite{veksler2018simulations} & 2018 & Cognitive modeling of users, attackers, defenders & ? \\
Ours & 2025 & [Insert focus] & 135 \\
\hline
\end{tabular}
\label{tab:other-reviews}
\end{table}
The use of simulation and modeling techniques in cybersecurity has been extensively studied in the literature (Table~\ref{tab:other-reviews}). Although several review studies have been conducted, most of them focus on specific application areas or modeling methodologies. 

In~\cite{jawad2021modeling}, studies of cyberattacks on~\acp{CPS} and industrial control systems are grouped by modeling methodology, considering approaches based on graphs, attack trees, and automata. In~\cite{chowdhury2021cyber} the authors reviewed cybersecurity research concerning the critical infrastructure sectors, with a particular emphasis on aviation, energy, and nuclear industries.

Other works focus on how simulations are applied to address different challenges in cybersecurity. In~\cite{kavak2021simulation}, five objectives of the simulations are considered: building representative environments (networks and connected systems), testing and evaluation (how simulations help evaluate the performance of cybersecurity tools), training (teaching basic security concepts to users, and how to recognize and react to a cyber threat to security professionals), risk analysis, and human factors in cybersecurity.

In~\cite{engstrom2022two}, the authors examine the evolution of cyber attack simulations over the past~20 years, highlighting trends, challenges, and methodological gaps. They classify simulations into three types: \emph{tactical} (replicating specific attack steps and responses), \emph{strategic} (analyzing long-term attack-defense dynamics and resource allocation), and \emph{impact} (evaluating consequences on cyber-physical systems and critical infrastructures).

Finally, in~\cite{veksler2018simulations} the authors analyze how simulations can incorporate cognitive models to enhance cybersecurity by focusing on human elements, such as the behaviors and decisions of attackers, defenders, and users. 


\section{Review Methodology}\label{sec:methodology}

\subsection{Paper Selection}

\begin{figure}[h]
\centering\includegraphics[width=\textwidth]{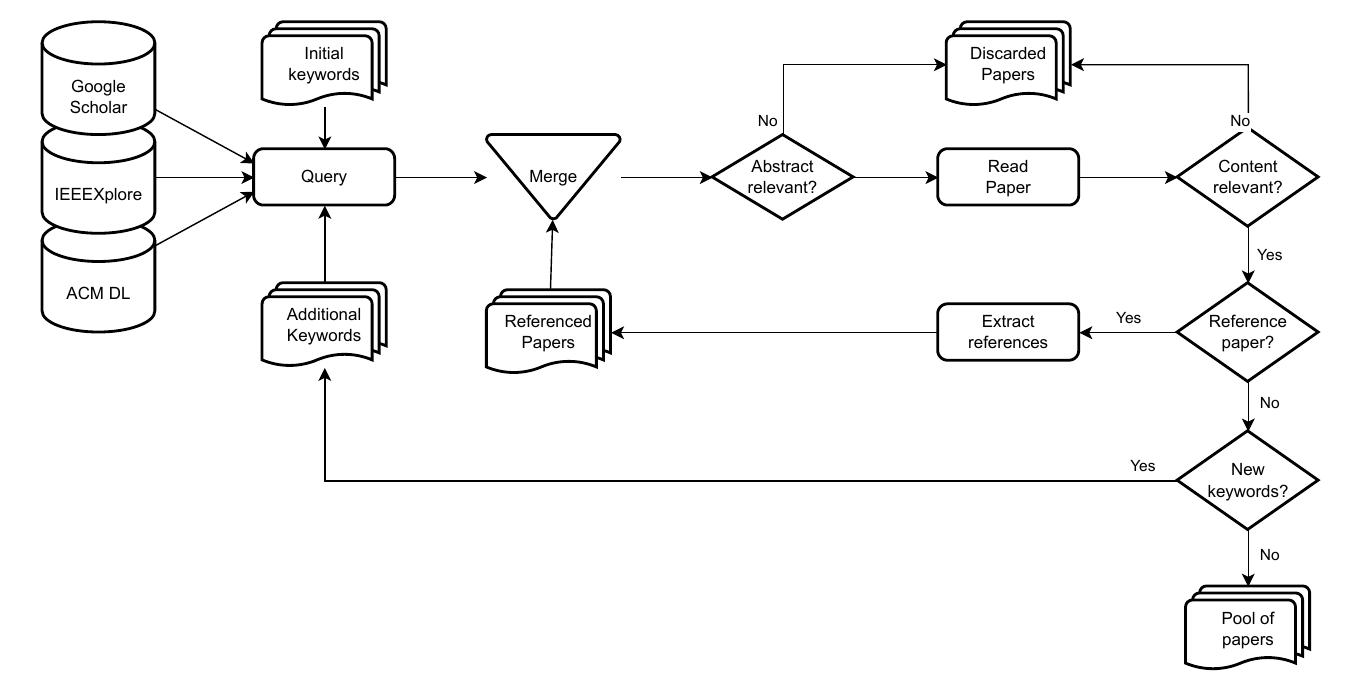}
\caption{Paper selection flowchart.}
\label{fig:selection}
\end{figure}

To ensure a comprehensive and consistent selection of research studies, review works need to define a rigorous methodology to retrieve the papers to examine~\cite{keele2007guidelines}, which is summarized in Figure~\ref{fig:selection}.

While initially some papers were included because they were known to the authors, the search process was conducted primarily using search engines like ACM Digital Library, IEEEXplore, and Google Scholar, considering for selection the first~$50$ results for each combination of keywords. 
The initial query was based on the following formula to capture a broad range of studies:
\begin{align}
(\textrm{cyberattack}~\textbf{or}~\textrm{cybersecurity})~\textbf{and}~(\textrm{modeling}~\textbf{or}~\textrm{simulation})\label{eq:query}
\end{align}

After the first round, additional terms that were frequently associated with simulation-based cybersecurity research have been identified, allowing us to extend the keywords for paper selection. As a result, new terms such as \textit{denial-of-service}, \textit{agent-based}, \textit{blockchain}, and \textit{malware} were fed to the bibliography search engines, enabling the inclusion of more specialized studies.
In addition to keyword-based searches, relevant survey papers were examined to retrieve further references that did not appear in the initial results, but were pertinent to the topic. 
Only papers that implemented actual simulation experiments were included, leaving out those that only offer methodological contribution. Furthermore, preprints, studies of insufficient quality, and duplicate entries were removed to ensure a consistent selection.

\subsection{Categorization and Analysis}

After excluding the papers that were not relevant, $135$ works were selected as part of the review: $91$~conference papers ($67.4\%$) and $44$ journal papers ($32.6\%$).

\begin{figure}[h]
\centering\includegraphics[width=.7\textwidth]{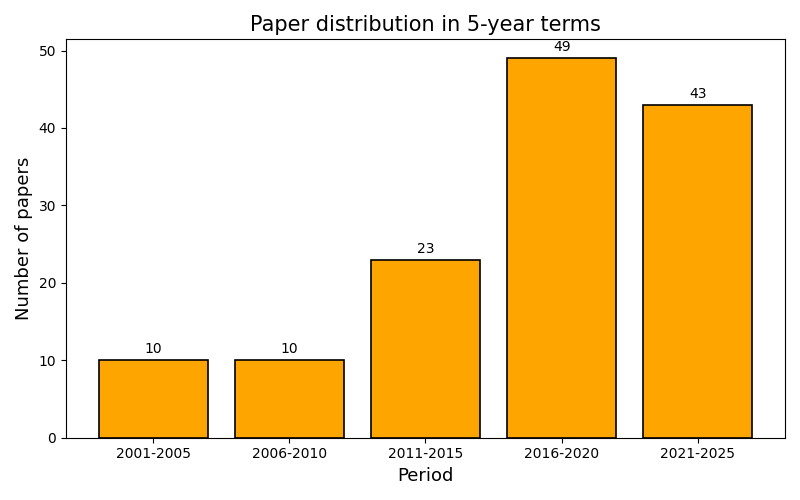}
\caption{Temporal distribution of the reviewed papers across five-year intervals over the past 25 years.}
\label{fig:years}
\end{figure}

Figure~\ref{fig:years} shows the number of papers by five-years term, revealing a growing interest in this type of study over the past ten years.

We also track the geographic areas of the authors’ home institutions, using a fractional counting method which works as follows: if a paper has~$n$ authors, then each one contributes~$1/n$ to the count of the geographic area of his/her home institution at the time the paper was written. In case of multiple affiliation, the first one is considered.

\begin{figure}[h]
\centering\includegraphics[width=.7\textwidth]{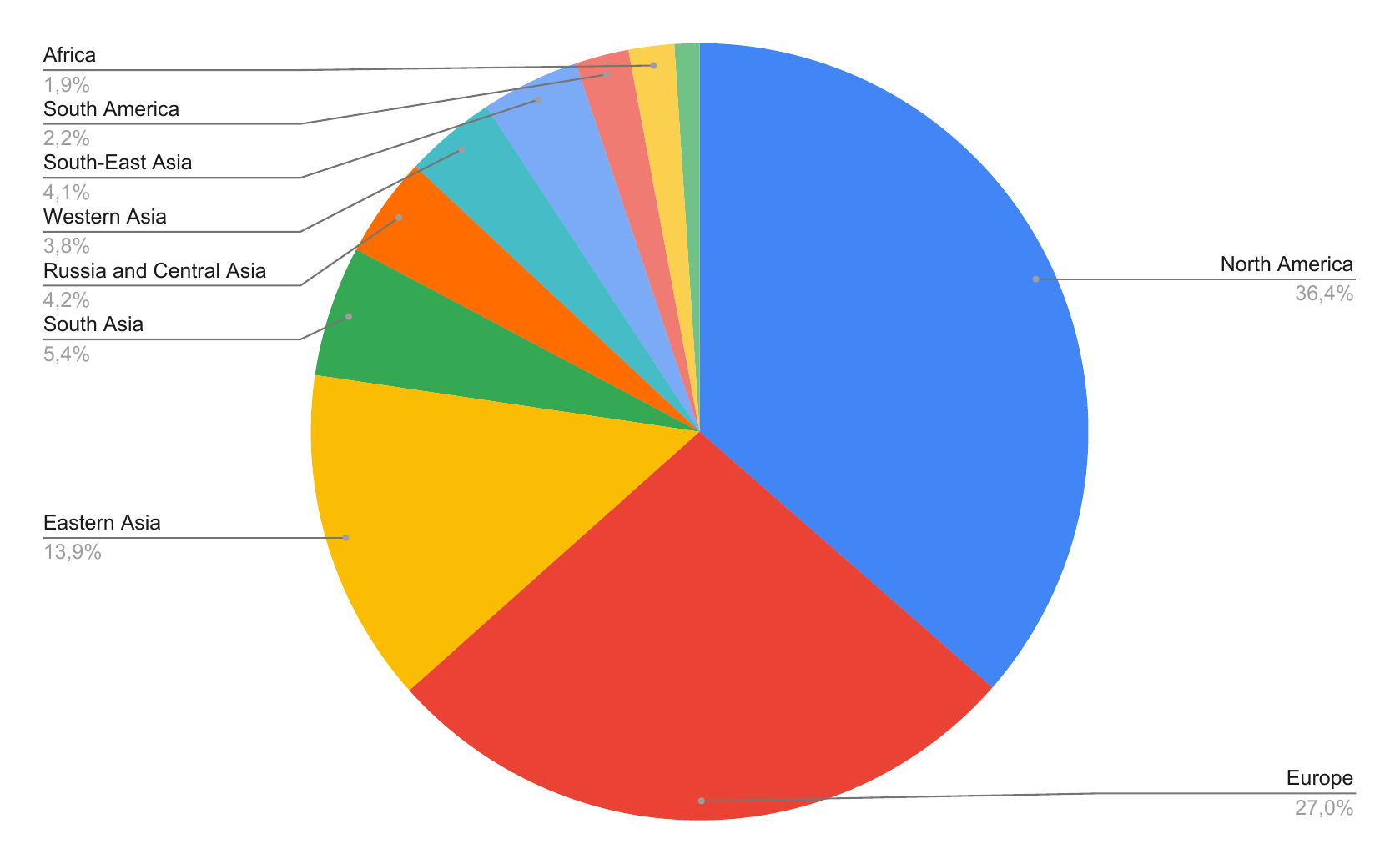}
\caption{Despite all continents are represented in this survey, more than $3/4$ of the papers are from North America (USA and Canada), Europe, or Eastern Asia countries (e.g., China, Japan, and Korea)}
\label{fig:areas}
\end{figure}

Figure~\ref{fig:areas} shows that the majority of papers are authored by researchers from North America, Europe, or East Asia, although every continent is represented by at least one publication. Specifically, the most recurrent nation is USA, followed by China, UK, Italy, India, and Canada.
On average, each paper has~$3.5$ authors, with a variance of~$2.49$. In~$23$ papers ($17\%$), different authors are affiliated with universities or research institutions from different countries, indicating international collaboration.

\section{Application domain}\label{sec:domain}

\begin{figure}[ht]
\centering\includegraphics[scale=\defaultscale]{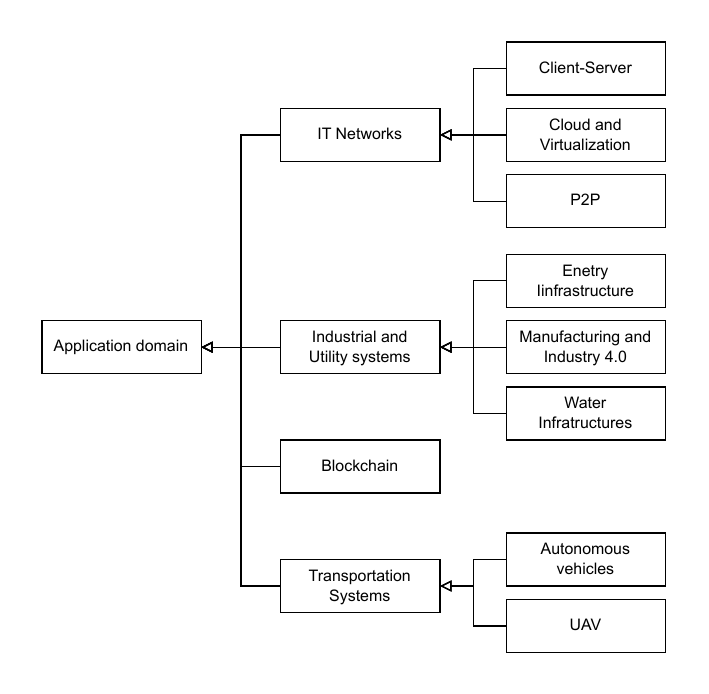}
\caption{Taxonomy of application domains.}
\label{taxonomy-domain}
\end{figure}

The analysis of the state of the art has revealed a significant diversification in the application domains of simulation in cybersecurity, as shown in Figure~\ref{taxonomy-domain}. The variety of areas covered reflects the growing awareness of the need for protection in both digital and physical systems.
In this section, we review some of the most investigated sectors, discussing how simulation is employed is such domains.

\subsection{IT Networks}
IT networks and infrastructures form the backbone of digital communications, and thus, they are important subjects for cybersecurity because they are the media through which attacks are conveyed. 
A critical challenge in this domain is mitigating \ac{DoS} attacks, which can severely impact network availability. 
In traditional enterprise networks, traffic routing and application of security policies are managed individually by switches and routers, whose behavior can be modified only with specific manual intervention.
On the other hand, \ac{SDN} centralizes network control, allowing for fast and coordinated responses to cyberattacks~\cite{zhao2022design}. Simulation can assess mitigation strategies, such as load balancing algorithms, evaluating their impact on packet loss, network stability, and server performance under attack conditions.

Client-server architectures are vulnerable to \ac{DoS} since the presence of centralized servers creates single points of failure that can be overwhelmed with a large volume of malicious traffic.
This scenario is analyzed in~\cite{kotenko2006simulation}, where the simulation environment is highly configurable, enabling the test of various attack taxonomies. Attack parameters include victim type (i.e., application, host, or network), \ac{DDoS} type, impact on the victim (i.e.,~ either disruptive or degrading), Agents’ set permanency (i.e.,~fixed or variable set of daemons), attack rate dynamics, exposure of malicious packets to detection filters, source address validity, and degree of automation. On the other hand, defense parameters include deployment location, mechanism of cooperation, covered defense stages, detection technique, and mitigation approach.

Beyond network availability, authentication is a crucial security aspect of IT infrastructures. In \cite{novak2017modeling}, interactions between users and authentication systems are modeled to evaluate the impact of memorization techniques like spaced repetition and system-generated passphrases. Cognitive burden metrics, such as Levenshtein distance and word count, are employed to measure the usability-security trade-off. Simulation assesses aggregate security against brute-force attacks, password reuse, and other vulnerabilities, showing the impact of password policies.
Authentication issues are also investigated in \cite{karagiannis2023cybersecurity}, in the context of medical imaging networks, together with unencrypted communications and access control weaknesses.

Military networks are a high-value target, as strategic infrastructures are more easily subject to attacks due to their geopolitical relevance.
In \cite{thompson2018agent}, \ac{ABM} is used to study malware propagation in mobile tactical networks that support military operations, incorporating hierarchical command structures, unit mobility, and short-range wireless communication. Military units are represented as agents, aiming to capture how group movement and hierarchical coordination influence risk exposure. 
Similarly, in \cite{dobson2017cyber}  cyber forces and networked environments are represented as interacting agents in a cyber warfare scenario. The proposed simulation mimics malicious operations like routing protocol manipulation, \ac{DoS}, and phishing attacks, while defensive forces deploy patching and mitigation strategies. 

Financial and enterprise networks are also appealing targets for attackers, given the valuable assets and sensitive information that they handle. Thus, protecting these infrastructures against cyber threats is crucial, as a successful breach would lead to financial losses and severe reputational damage.
In \cite{daah2025simulation}, the authors examined a scenario where a Zero Trust Architecture (i.e.,~no user device can be trusted by default) is integrated with a hybrid access control system and blockchain technology. Simulation helps assessing the impact of cyberthreats like \ac{DoS}, \ac{MitM}, zero-day exploits (i.e.~vulnerabilities unknown to developers and thus lacking countermeasures), and smart contract vulnerabilities, using metrics such as detection accuracy, false positives/negatives, and response time. 
In \cite{moskal2018cyber}, an adversary model defines the strategy to penetrate an enterprise network based on \textit{intent}, \textit{capability} (i.e., the skillset of the attacker), \textit{opportunity} (i.e., possible actions based on
the attacker's intent and accumulated knowledge), and \textit{preferences} of the attacker. 

While in client-server architectures \ac{DoS} typically targets a centralized server, in \ac{P2P} systems resources are distributed among the various participants, making it harder for an attacker to prevent access to online resources. However, in~\cite{qwasmi2011simulation} it was demonstrated that attacks targeting an individual node can negatively affect the whole network, slowing down system performance and undermining the availability of the service.

Finally, due to their increasing popularity, cyber risks to cloud and grid systems must be taken into account. In \cite{vasenin2014environment}, the authors simulated virus propagation in an IaaS infrastructure, showing cascading effects, and \ac{DDoS} targeting multiple types of cloud systems. The experiments also evaluate the role of \acp{IDS} to mitigate the impact of these threats.

\subsection{Industrial and utility systems}
Protecting industrial infrastructures and utility systems from cyberattacks is of paramount importance, as the impairment of these systems could have widespread social and economic consequences. The digitalization process driven by the Industry 4.0 paradigm and the growing interconnection of modern industrial networks has exposed these environments to further cyber risks. 
Simulations in this field may need to take into account multiple factors, such as the behavior of the \ac{CPS} and the organization of internal communication networks, allowing researchers to examine the impact of cyberattacks, such as false data injection and \ac{DoS}~\cite{le2020gridattacksim}.
Attacks on \acp{CPS} often take advantage of specific properties of industrial control environments, such as the modification attack on smart grids, where adversaries manipulate measurement data to deceive control systems. Simulation helps to assess how injected biases into the control signals can destabilize the system under various scenarios, affecting both angle and voltage stability~\cite{chen2013impact}. 
\ac{MitM} and data injection attacks, as discussed in \cite{stefanov2014cyber}, can significantly alter system behavior, leading to incorrect operator decisions. Similarly, in \cite{stuanculescu2021case}, the authors have examined the effects of induced faults on transformers within a petrochemical plant. 
The way that \ac{DoS} can disrupt sensor data collection and control instruction transmission is analyzed in \cite{ding2017attacks}, showing that attack-induced delay decreases the stability of the system.
The complexity of industrial environments is further influenced by the interaction between human operators and automated control mechanisms. 

In \cite{cunningham2017adapting}, \ac{ABM} is used to simulate security failures in power grid operations, considering as a use case the Northeast Blackout of 2003, and taking into account people, tools, and organizations involved. While the blackout itself was not caused by malicious actors, it is shown that its key failures, such as system misconfigurations, communication breakdowns, and software bugs, could realistically be replicated by cyberattacks.

In \cite{huang2018case}, a scenario where attackers manipulate SCADA systems to open circuit breakers was modeled based on Ukraine power system hacking events, evaluating system vulnerabilities and the risk of large-scale blackouts.

\subsection{Transportation Systems}
The growing popularity of IoT has also extended to the automotive industry, where cybersecurity is an important challenge for connected and autonomous vehicles. In fact, weaknesses in Vehicle-to-Everything communications could be exploited to modify cars' behavior, dangerously undermining road safety. 
In this field, simulation is used to assess vulnerabilities in transport networks and potentially prevent attacks that could compromise road safety. For instance, in drive-by downloads attacks the navigation systems are manipulated, leading vehicles to ignore traffic signals, make unintended turns, or collide with other vehicles and pedestrians~\cite{malik2020analysis}.
For these investigations, multilevel modeling is frequently used to incorporate traffic simulation, network simulation, and potentially other relevant aspects, such as 5G control place~\cite{santos2023towards}.
As in nearly every sector, a significant threat to vehicular networks is \ac{DoS}, where malicious actors overwhelm the network with a huge amount of messages that must be processed, causing delays or failures in critical safety messages.
To mitigate these concerns, in \cite{verma2013prevention} the authors tested the effectiveness of Bloom Filter detection, a method that employs a space-efficient probabilistic data structure to identify anomalies in packet transmission. 
Furthermore, the increasing integration of intelligent devices in transport requires studies to ensure the reliability of communication networks and data protection, considering the interconnections among the involved actors.
In \cite{jackson2023agent}, \ac{ABM} is used to represent vehicles, pedestrians, and roadside infrastructure as distinct agent populations that interact via communication channels such as vehicle-to-vehicle and vehicle-to-infrastructure. The goal of the tests is to investigate the propagation of attacks originating from different agents, considering parameters such as infection probability, detection, and defense capabilities. 
In \cite{cassou2020simulation}, the authors investigated the impact of jamming, replay, falsification, and congestion attacks on ITS-G5 vehicular communication systems. Each attack exploits known weaknesses: jamming interferes with wireless communication by transmitting noise packets on the same frequency, replay attacks resend outdated packets destabilizing vehicle coordination, falsification attacks modify critical parameters of transmitted messages, and congestion attacks overwhelm communication channels.
Besides ground transportation, aircraft safety - in particular in the context of \ac{UAV} networks - has also been object of research, with studies examining cyber threats like \ac{DDoS}, jamming attacks~\cite{javaid2013uavsim} (i.e.,~deliberate interference with wireless communication), and data tampering~\cite{puchaty2011performance}. 

\subsection{Blockchain}\label{sec:blockchain}
Finally, certain types of attacks are inherently linked with the type of system under examination, as they exploit protocols or technological features. In blockchains, cyberattacks often exploit the very features of the technology, due to its open and decentralized nature. In particular, blockchains are vulnerable to attacks targeting consensus protocols, smart contracts, and the network structure. 
Some known threats that have been investigated in scientific documents are:
\begin{itemize}
    \item \textit{51\% attack}, which occurs when one party controls more than half of the system’s computing power, potentially enabling transactions manipulation and undermining the decentralized nature of the system. In fact, some blockchains are based on Proof-of-Work, a consensus mechanism where participants must solve complex cryptographic puzzles to generate new valid blocks. Attackers can exploit their computational power to mine a large number of blocks and carry out fraudulent behavior, such as double spending~\cite{ye2018analysis}.
    \item \textit{Sybil attack}, where a malicious user creates a large number of fake identities (Sybil nodes) in the P2P network, in order to obstruct regular message dissemination~\cite{tomar2024blockchain}. While often associated with blockchain networks, Sybil attacks can also target IoT systems, wireless sensor networks, and other distributed environments~\cite{rajan2017sybil}. 
    \item \textit{Stalker attacks}, where adversaries attempt to exclude blocks of targeted miners from the main chain~\cite{chicarino2020detection}.
    \item \textit{Saving attack}, where a malicious validator in Proof-of-Stake blockchains (i.e.~, block validators are selected based on their stake instead of computational work) withholds its right to propose blocks during a temporary consensus failure and later use them to provoke other consensus failure~\cite{otsuki2021impact}.
    \item \textit{Denial of Chain}, where attackers manipulate consensus mechanisms to enforce an alternative chain as the main chain, by withholding or selectively publishing blocks, then overriding legitimate transactions~\cite{bordel2021denial}.
    \item \textit{Re-entrancy attacks}, where vulnerabilities in Ethereum smart contract execution flow are exploited to repeatedly withdraw funds before state changes are committed~\cite{moubarak2018blockchain}.
    \item \textit{Sandwich attack}, where an attacker exploits the visibility of pending transactions in the mempool. The attacker places two transactions around a victim's trade — one before and one after — to manipulate the asset price and extract a profit at the victim's expense~\cite{stucke2022simulation}.
    \item \textit{Partitioning attack}, where an adversary splits the network into isolated subgroups, preventing nodes in different partitions from communicating~\cite{wuthier2021proof}.
    \item \textit{Transaction malleability}, where the ID of a transaction is manipulated before it gets inserted on the blockchain, allowing the attacker to modify and resend a transaction at a later time, potentially leading to double spending or network inconsistencies. The attack is particularly dangerous in the context of e-voting, where network delay and block generation rate are impactful parameters~\cite{khan2020simulation}.
\end{itemize}


\subsection{Discussion}
\begin{table}[ht]
\centering
\caption{Application domains and key simulation characteristics}
\begin{tabular}{|l|c|c|}
\hline
\textbf{Domain} & \textbf{Common Threats} & \textbf{Modeling Focus} \\
\hline
IT Networks & \makecell{ DoS \\ Authentication Issues\\ MitM} & \makecell{Network Resilience \\ Packet Loss \\ Access Control} \\
\hline
Industry \& Utility &  \makecell{DoS \\ MitM \\Data  Injection} & \makecell{Operability \\ SCADA Security} \\
\hline
Transportation & \makecell{DoS \\False Data Injection} & \makecell{Road safety \\ Vehicular Communication} \\
\hline
Blockchain & \makecell{Consensus Attacks \\ Smart Contract Bugs} & \makecell{Double Spending \\ Consensus Failure} \\
\hline
\end{tabular}
\label{tab:domain}
\end{table}

While certain cyber threats are cross-class issues, others are characterized by domain-specific vulnerabilities, and thus need to be modeled accordingly as summarized in Table~\ref{tab:domain}.
The level of system exposure to external networks is pivotal in establishing the pathway for an attack. The increasing digitalization of services, adoption of cloud technologies, remote access, and real-time monitoring have significantly increased the attack surface of many systems, providing attackers numerous entry points that can potentially be exploited to perform malicious actions.
While isolated environments may still be susceptible to cyber threats, such as attacks launched by internal actors, the limited engagement with the external world reduces the likelihood of remote attacks.

After successfully penetrating a system, the way attacks propagate is highly contingent upon the nature of the domain and the goal of the adversaries. Attackers tend to move laterally, gaining control of the system incrementally, until the ultimate target is reached.

\section{Cyber Threats}\label{sec:threats}

\begin{figure}[h]
\centering\includegraphics[scale=\defaultscale]{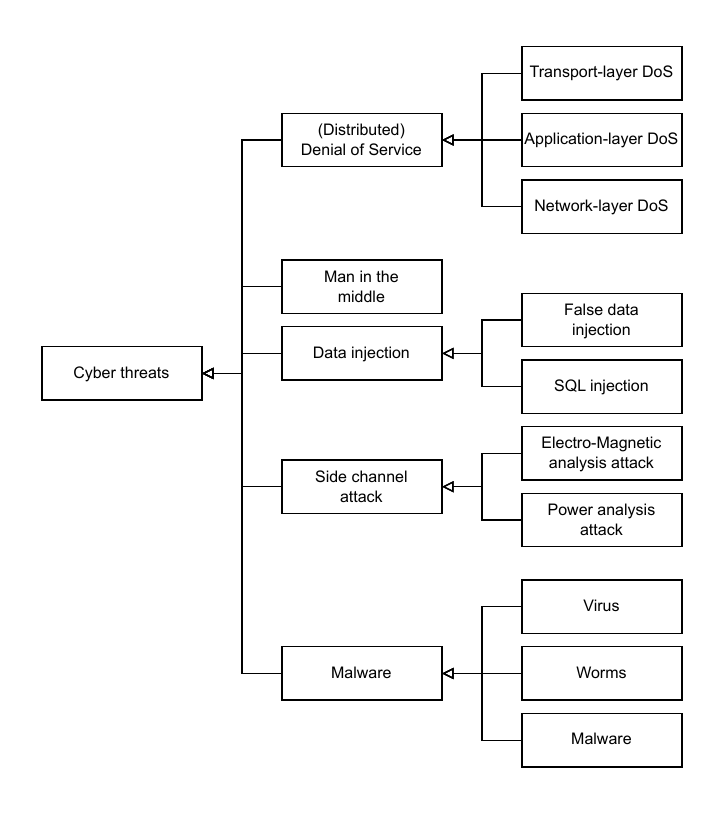}
\caption{Taxonomy of cyber threats.}
\label{taxonomy-threats}
\end{figure}

Several cyber threats exist, differing in targets, goals, attack strategies, and vulnerabilities exploited as summarized in Figure~\ref{taxonomy-threats}. 
Some attacks rely on overwhelming network resources, while others exploit software vulnerabilities, manipulate user behavior, or operate stealthily over extended periods.
The nature of each threat determines which modeling approaches can be employed, the complexity of the representation, the metrics being used, and the simulation parameters.
In this section, we discuss the challenges linked with the most investigated cyber threats.

\subsection{Denial of Service}
Out of all the cyber threats, \ac{DoS} is the most extensively investigated through simulation. This is not only due to its impact on cybersecurity, but also because these attacks perfectly match with a simulation approach, as modelers can represent both attackers and defending entities as agents that interact dynamically within a simulated network environment~\cite{kotenko2010agent}. 
DoS serves as an umbrella term and generally applies to any attempts to disrupt the normal operation of a system or service by overwhelming it with a volume of requests that the system cannot handle, rendering it unavailable to legitimate users. The effectiveness of \ac{DoS} can be enhanced when multiple compromised devices are working together, which are often part of a botnet that overwhelms the target system with a huge amount of traffic. DDoS may be particularly difficult to defend, since the presence of several devices utilized to conduct the attack makes it harder to filter out the malicious traffic without interfering with the legitimate users. As summarized in table~\ref{tab:dos}, \ac{DoS} attacks can be conducted against multiple layers of the Internet stack, with different implications for system availability.
At the MAC layer, attackers can take advantage of the capture effect and unfairness characteristics found in the IEEE 802.11 protocol to dominate channel access~\cite{gupta2002denial}. At the network layer, flooding attacks overwhelm bandwidth with an excessive transmission of IP packets, saturating the bandwidth of communication channels. At the transport layer, connection-handling mechanisms are exploited to exhaust server resources, leading to unresponsiveness. Finally, at the application layer, attackers can overload specific protocols, such as flooding IEC 60870-5-104 packets in SCADA systems~\cite{kalluri2016simulation}, to delay time-critical processes, and even shutting down normal operations.

DoS employes different techniques, depending on how the targeted network is configured, the type and volume of the generated traffic, the protocol weaknesses being exploited, and the attack's distribution across single or multiple sources. Several studies have examined DoS attacks using one or more of the following techniques:
\begin{itemize}
    \item \textit{Slowloris} where a targeted web server’s connections are exhausted by sending incomplete HTTP requests at a slow rate~\cite{sabri2021slowloris}.
    \item \textit{IP/MAC Spoofing}, where the attacker forges its network data in order to disguise malicious traffic and evade detection mechanisms~\cite{oktian2014mitigating}.
    \item \textit{Bulky Message}, where a target server is flooded with excessively large messages or files.
    \item \textit{Garbage Message}, where a high volume of meaningless packets forces network devices to fill up the switch’s flow table.
    \item \textit{SYN flood}, where a large number of SYN requests are sent to a target server without completing the connection\cite{bogdanoski2013analysis, zhang2017opnet, lian2007simulation}.
    \item \textit{UDP storm}, where multiple hosts send a large volume of UDP packets to random ports of the target machine, causing it to repeatedly check for listening applications and respond with ICMP Destination Unreachable messages~\cite{asri2015impact}. 
    \item \textit{Ping of Death}, where a host transmits numerous ICMP Echo Requests with an oversized or malformed packet size to a target system~\cite{ ni2018cyber}.
    \item \textit{RTS/CTS}, where an attacker exploits the Request-to-Send and Clear-to-Send mechanism in IEEE 802.11 networks by sending forged RTS/CTS frames with excessively long duration values, thus reserving the wireless channel for an extended period, preventing legitimate devices from accessing the medium~\cite{nagarjun2013simulation}.
    \item \textit{Disassociation on 802.11 wireless networks}, where an attacker spoofs a legitimate access point or client and sends fake disassociation frames to disconnect devices from the network, forcing victims to repeatedly reconnect~\cite{aslam2008802}.
    \item \textit{Process Table attack}, the target system’s process table is filled with excessive requests, preventing new processes from being created and rendering the system unresponsive until the attack stops or the administrator manually terminates the malicious processes~\cite{razak2002network}.
    \item \textit{Mailbomb}, where a mail server is overwhelmed by sending a massive number of emails from multiple sources, leading to excessive resource consumption and disruption of email services~\cite{shin2023beyond}.
    \item \textit{HELLO Flood}, where HELLO messages, used to announce the presence of a node, are broadcast to trick other nodes into incorrectly recognizing the attacker as a legitimate neighbor~\cite{lotfy2013performance}.
    \item \textit{DNS Flood}, where the attacker exploits open DNS resolvers by sending small queries with a spoofed IP address, causing the servers to send large responses to the victim~\cite{furfaro2015simulation}.
\end{itemize}

\begin{table}[ht]
\caption{Classification of DoS Techniques by OSI Layer}
\begin{tabular}{ll}
\toprule
\textbf{OSI Layer} & \textbf{DoS Attack Technique} \\
\midrule
MAC (Layer 2) & RTS/CTS Attack \\
              & Disassociation Attack \\
\hline
Network (Layer 3) & IP/MAC Spoofing \\
                  & UDP Storm \\
                  & Ping of Death \\
                  & HELLO Flood \\
\hline
Transport (Layer 4) & SYN Flood \\
\hline
Application (Layer 7) & Slowloris \\
                      & Mailbomb \\
                      & Bulky Message Flood \\
                      & Garbage Message Flood \\
                      & DNS Flood \\
\hline
System-level & Process Table Attack \\
\bottomrule
\end{tabular}
\label{tab:dos}
\end{table}

\subsection{Man-in-the-Middle}
\ac{MitM} is a severe threat to confidentiality and integrity of data transmission, particularly in scenarios where communication channels are encrypted, authentication mechanisms are weak, or network access is poorly controlled. 
While simulation might not be the most appropriate approach to evaluate the likelihood of data leakage, it is well-suited to evaluate the consequences that insecure communication could have on system operations, above all in case the attacker is able to inject harmful data.

In \cite{yan2011cyber}, simulation is used to
investigated how a \ac{MitM} can affect the behavior of the power system of a wind farm, leading to overspeed conditions, voltage instability, and potential equipment damage. The \ac{MitM} targets the optical fiber link between the control center and the wind turbines, intercepting and modifying measurement data and control commands. As a result, operators are led to think that the system is functioning regularly, despite turbine operations are destabilized. 
A similar attack scenario is examined in \cite{choi2020multi}, where an adversary is able to eavesdrop on the network backbone of a power distribution automation system. The attacker injects false measurements regarding the status of electrical components, to mislead operators in their decision-making, such as switching when it is unnecessary or failing to respond to a legitimate fault response. However, it was shown how mitigation strategies based on shared secret keys and message authentication codes enable the detection of tampered data.
Furthermore, to increase the veracity of the representation, modelers could feed simulators with real network traffic, ensuring that packet structures and communication patterns reflect actual system behavior. For instance, in \cite{fritz2019simulation}, Wireshark is employed to capture and analyze live network traffic between components of a smart grid.

Finally, in \cite{abdo2024vehicular} a \ac{MitM} creates a large number of pseudonymous identities and threatens road safety by falsifying data exchanged among connected vehicles. These data contain information such as vehicle size, position, speed, heading, acceleration, and brake system status, which could lead cars to make unsafe driving decisions. The experiments show that if a certificate mechanism is employed, vehicles can swiftly block forged messages.

\subsection{Data Injection} 
Data injection occurs when attackers feed malicious code into a target system, exploiting vulnerabilities in input validation and data handling, with the aim of executing fraudulent operations that may cause data leaks or unauthorized access.
Code injection, and especially SQL injection, has been a well-documented cybersecurity threat for decades, consistently ranked among the OWASP Top 10 vulnerabilities~\cite{patil2023review} due to its frequency and potential effects. 
Despite its impact on cybersecurity, simulation-based investigations remain quite uncommon. The primary reason is the difficulty in parameterizing data injection attacks. Unlike \ac{DoS}, which can be studied observing how system degradation is influenced by varying traffic loads, data injection success outcome is binary, making it unsuitable for quantitative analysis. Furthermore, the success of the attack usually depends on application-specific factors, such as database structure, query management, and input handling.
As a result, penetration testing and code analysis are preferred for assessing injection risks.

Despite these limitations, few simulation studies have been conducted, in particular to analyze the consequences of undetected data injections on the targeted system. 
In the context of vehicle-to-everything communication, false data regarding the surrounding vehicles or the driving environment may lead to the loss of car control, endangering drivers, passengers and pedestrians, and disrupting the traffic flow~\cite{folan2023cybersecurity}. In SCADA systems, the injection of falsified measurements can deceive control decisions, such as triggering unnecessary disconnection of transmission lines or altering generator outputs, potentially destabilizing the power grid~\cite{lingaraju2021simulation}.
Also in shipboard power systems, false data injection is concerning issue, as attackers can manipulate load control instructions to deliberately reduce the electricity supply by forcing incorrect power adjustments.
Specifically, in \cite{kushal2018risk}, the attack consists of injecting false commands into the control system, which unknowingly transmits the incorrect instructions to the generators, resulting in a distribution system disturbance. The system's response is assessed by measuring the deviation from generation and estimating load curtailment. To defend from the attack, an independent control mechanism is considered to detect and block the modified commands to maintain the stability of the system, effective demand response, and manage energy supply.
In \cite{tobin2015simulating}, the authors simulated a SQL injection, where fraudulent SQL code interferes with database queries. 
The model represents communication of different network components within a typical demilitarized zone architecture. Network traffic generated during the simulated attack was captured and analyzed using Wireshark to observe SQL injection attempts and examine the interactions between the attacker and the server.
In \cite{blancaflor2024strengthening}, SQL injection is examined through a penetration test conducted within a controlled simulation environment. The SQL injection attack is executed on a locally hosted demo website, starting with reconnaissance to identify vulnerabilities in authentication mechanisms. The simulation allows for testing automated attack techniques that exploit input validation weaknesses to gain unauthorized access, while the penetration test assesses how different exploitation techniques affect data confidentiality. Phishing is also studied, demonstrating how fraudulent login prompts can be used to deceive users into revealing credentials.

\subsection{Malware}
Malware refers to any type of malicious software designed to cause harmful or unintended behavior when executed on a device, usually with the aim to cause damage, steal sensitive information, extort money, or conduct espionage. Certain malware, like viruses and worms, do not only produce local effects, but attempt to propagate to reach other devices.
While sandboxed environments are preferable for evaluating the effects of malware on infected devices, simulation is often used to study dissemination dynamics.

In \cite{leszczyna2010simulating}, viruses and worms are represented as autonomous agents that replicate and propagate across the information system of a power plant. Once they reach the target, they deactivate the hosts’ network cards, rendering unavailable control and remote monitoring services. The simulation helps investigate both the propagation dynamics and the operational implications of the attack.
In \cite{hosseini2016agent}, network nodes are modeled as autonomous agents that interact within a scale-free network. Each agent transitions between states following a SEIRS model: \textit{susceptible} agents are vulnerable to infection, \textit{exposed} agents carry inactive malware, \textit{infectious} agents spread the malware to their neighbors, and \textit{stifler} agents stop transmitting the infection but may become susceptible again. The model also incorporates defensive strategies such as selectively immunizing the nodes with the highest connectivity and software diversity, where automatic program transformations are applied to generate variants of applications.
Similarly, in \cite{benomar2022agent}, 5G mobile devices act as mobile agents in an urban environment structured as a Poisson-Voronoi tessellation, a mathematical model that represents city street layouts with random segmentation. The simulation employs an SI model, where an initially infected device transmits malware to susceptible neighbors if they remain within communication range for a sufficient time. 
In \cite{hara2024extending}, the NS-3 simulator is extended to analyze Mirai, a self-propagating botnet malware that targets \ac{IoT} devices to perform \ac{DDoS} attacks. Captured traffic is forwarded through a router that intercepts communications and redirects packets based on protocol and port numbers. 
Mirai was further investigated in \cite{tanaka2017modeling}, where its infection behavior was modeled using Petri Nets alongside Hajime, a peer-to-peer IoT malware that spreads by exploiting vulnerabilities, blocks ports to prevent reinfection, and vanishes upon reboot. Both IoT devices and malware are represented as agents, with transitions defining infection events. Different reboot frequencies and network topologies are tested to analyze infection dynamics, showing their impact on malware dissemination and how interactions between Mirai and Hajime affect infection patterns.

\subsection{Side-Channel Attacks}

Side-channel attacks exploit unintended data leakages that occur during computing operations, usually related to cryptography. The collected physical signals, such as energy consumption, execution time, or electromagnetic emissions, are used to infer information like a secret key.
To analyze the feasibility of these attacks, it is first necessary to reproduce realistic power traces.
In \cite{bhasin2013physical}, two methods are compared. One uses detailed electrical simulations to estimate how the chip's power consumption changes during encryption, taking into account the effects of wiring and circuit structure. The other relies on faster digital simulations, where the activity of the circuit is monitored and power consumption is approximated by counting signal changes.
Small variation in power usage during the encryption procedure can reveal information about the secret keys.
In \cite{kumar2017efficient}, the authors investigate how much information leaks from a chip through electromagnetic signals, focusing on the AES encryption algorithm. The simulator first mimics the circuit’s activity to determine current behavior, then uses currents on the chip’s top metallization layers to estimate electromagnetic emissions, which in turn are employed to generate virtual signals that a nearby electromagnetic probe might capture. By repeating this process for many different encryption runs and analyzing the resulting signals, the model assesses whether an attacker could recover the full secret key.
A similar approach is used in \cite{menichelli2008high}, where both hardware and software implementations of AES are simulated at a high level with a tool based on SystemC. The internal values of registers and memory during encryption are tracked, enabling the estimation of power usage based on how those values change over time. This lays the foundation to generate power traces that resemble what would be measured on a physical chip. 


\subsection{Discussion}
\begin{table}[ht]
\centering
\caption{Cyber threats and simulation suitability}
\begin{tabular}{p{.3\textwidth}p{.3\textwidth}p{.3\textwidth}}
\toprule
\textbf{Threat} & \textbf{Typical Targets} & \textbf{Simulation Goal} \\
\midrule
DoS / DDoS & Web servers, Network devices & Measure impact through tunable parameters \\
\midrule
MitM & Communication channels that lack proper security measures & Evaluate consequences of insecure communication \\
\midrule
Data Injection & SCADA system, Web apps, Connected vehicles & Study the consequences of undetected injections on the system \\
\midrule
Malware & Personal devices Critical infrastructures & Study malware propagation \\
\midrule
Side-Channel & Devices performing cryptographic operations & Assess the feasibility of the attack \\
\bottomrule
\end{tabular}
\label{tab:threats}
\end{table}

Cybersecurity includes a wide range of threats, differing in the objectives, exploited vulnerabilities, and modes of propagation. Although not all cyberattacks are equally suited for detailed modeling, simulation can still serve different purposes depending on the type of threat, as summarized in Table~\ref{tab:threats}. Simulation is well-suited when it is feasible to reproduce realistically the sequence of actions of the involved parties and their effect on the system, and when the impact of the attacks is measurable with meaningful metrics.

The cyber threat that is best suited for simulation is \ac{DoS}, as the effects of the attack depend on easily configurable parameters such as traffic volume, number of attacking nodes, and target system capacity. These parameters allow researchers to test different attack intensities and to evaluate how network performances degrade over time, defining threshold conditions. Additionally, \ac{DoS} simulations produce measurable outcomes, such as increased latency, packet loss, and server unresponsiveness, allowing the comparison of different mitigation strategies. 
In contrast, the outcome of certain attacks depends primarily on the exploitation of vulnerabilities that, being unknown, cannot be analyzed in advance. Also, when success is determined entirely by system configuration, static analysis or penetration testing are more appropriate approaches. 
Similarly, attacks that depend on deception, such as phishing or social engineering, are difficult to model in a simulation environment since their effectiveness relies on human behavior rather than predictable system responses.

\section{Modeling Technique}\label{sec:modeling}

\begin{figure}[h]
\centering\includegraphics[scale=\defaultscale]{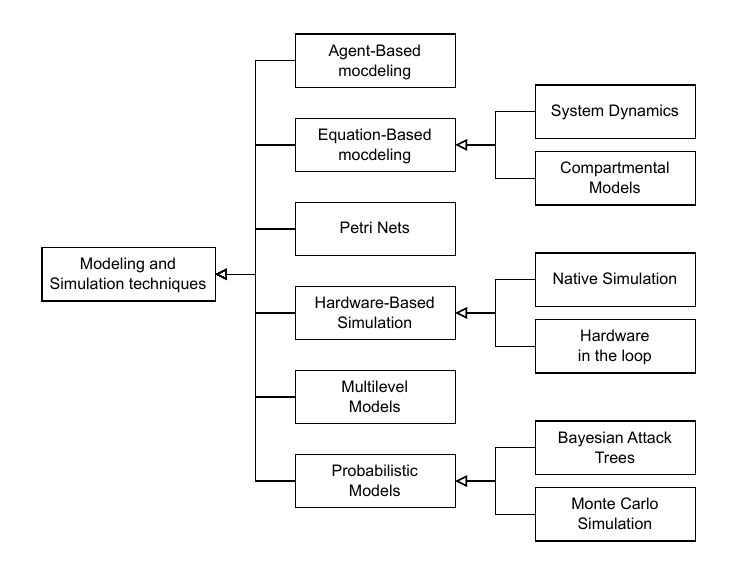}
\caption{Taxonomy of modeling approaches.}
\label{taxonomy-modeling}
\end{figure}

Due to the diversity of cyber threats and application domains involved, there are currently no methodological standards for cybersecurity simulation. As a result, researchers select modeling approaches based on the specific features of the threat, the complexity of the environment, and analysis objectives.
In this section, we introduce the most popular modeling paradigms (summarized in Figure~\ref{taxonomy-modeling}, discussing the context where they are more suited.

\subsection{Agent-Based Modeling}
\ac{ABM} has a prominent role in cyberattack simulation thanks to its ability to represent in detail the behavior of all entities involved in the scenario under investigation.
In~\cite{koutiva2021agent}, both attackers and targets are represented as agents within a critical water infrastructure. Attackers are classified by expertise level (amateurs, experts, and highly skilled adversaries), which defines the behavioral rules in the attempt to locate and exploit system vulnerabilities. Adversarial actions are dictated by resource availability, estimated attack success probability, and strategies of defenders. On the other hand, targets include cyber-physical components such as sensors, actuators, PLCs, and SCADA components, each with different levels of protection. Agents' behavioral rules can be either predefined based on expert knowledge or derived from real-world samples, such as malware execution traces or attack logs. 
In~\cite{furfaro2017using}, agents are defined using state machines, where transitions between states are determined by analyzing logs from infected systems through process mining techniques. 
In~\cite{kotenko2005agent}, \ac{DDoS} agents operate in a coordinated way: a master issues commands to daemons, which execute the attack by flooding the target with malicious traffic. Instead, defensive agents monitor network traffic, detect anomalies, discard malicious requests, and attempt to trace attack origins, adapting their countermeasures in response to threats. 
In~\cite{poisel2013game}, the modeling DDoS attacks on critical infrastructures incorporates game theory aspects, as attackers and defenders engage in anticipation games. The various agents predict the opponent’s next move to optimize their actions, using a dependency graph to represent the relationships between infrastructure components, attack paths, and defensive responses. Each agent is characterized by a set of attributes and decision rules that defines its behavior. Attackers analyze vulnerabilities and launch botnet-based assaults, while defenders implement countermeasures based on available resources and information from system monitoring. 

While most agent-based models in cybersecurity focus on \ac{DoS} attacks, other cyber threats can also be analyzed through \ac{ABM}. In~\cite{renaud2013simpass}, the authors developed a simulator that evaluates the effectiveness of password policies, quantifying outcomes such as login success rates, system breaches, and password leakages under different security rules. The model defines four agent types. \textit{Employees} authenticate across internal and external systems, sometimes forgetting, reusing, or sharing passwords. \textit{Malicious insiders} exploit these behaviors for unauthorized access, while \textit{hackers} use brute-force attacks, default passwords, or stolen credentials to infiltrate. Finally, \textit{system administrators} enforce policies but may neglect maintenance, increasing vulnerabilities.

While most of the models focus on attackers and defensive mechanisms, the quality of the analysis can benefit from incorporating human behavior and decision-making, as user actions play a crucial role in system security.
To accomplish this, in~\cite{rausch2018modeling} the authors introduced the \ac{GAMES} formalism, where decision making of users, defenders, and attackers is defined by an algorithm that moves each agent toward its most advantageous state. By simulating human actions, errors, and adaptations, \ac{GAMES} provides a more realistic assessment of risk, system vulnerabilities, and defense effectiveness, offering insights into how human factors influence cybersecurity resilience~\cite{rausch2018modeling}.
In \cite{chiong2008modelling}, user agents within a database system are assigned with specific privileges and access rules in order to comply with the discretionary access control policy, where permissions are granted selectively. Agents operate hierarchically, determining access paths and enforcing security policies. Each agent maintains a reliability index, which influences future access privileges and can lead to restrictions if data corruption is detected.

Finally, \ac{ABM} allows multiple levels of detail to coexist. For instance, in \cite{kotenko2003agent}, macroscopic modeling is used to capture high-level attack strategies and their overall impact on the system, while microscopic modeling focuses on individual attack steps and their execution details. 

\subsection{Equation-based Modeling}
\ac{EBM} enables a system-wide representation of complex behaviors using mathematical equations, making it a valuable alternative to \ac{DES} and \ac{ABM}. While ABM is able to represent the heterogeneity of the behavior of the various entities involved with a high level of detail, its execution can be computationally expensive and impractical for large-scale, long-term analyses.

In particular, \ac{SD} is a modeling methodology where the behavior of the system is described by differential equations that integrate feedback loops to capture interactions among system components. Two main entities are used to describe the system: \textit{stocks} representing cumulative elements and \textit{flows} representing the rates of change of the stocks. This approach can be used to model the evolution of cyberattack conditions, and it is especially well suited for analyzing the attack effects in complex interconnected systems like university IT infrastructure~\cite{kannan2016modeling}. 
In \cite{genge2015system}, \ac{SD} is employed to evaluate the impact of cyberattacks on critical infrastructures, modeling interactions between physical and cyber layers. In particular, sensitivity analysis is employed to compute the effects of cyberattacks on control variables and their propagation throughout the system. 
In \cite{medoh2022future}, \ac{SD} assesses how cybersecurity initiatives impact business resilience in Industry 4.0, capturing the interdependencies between security investments, operations, and risk mitigation. The model enables what-if scenario analysis to evaluate long-term cybersecurity decisions. The study gathers data from 150 professionals in network and security organizations to identify key parameters like resource allocation, employee awareness, and time management. These inputs shape the SD framework, which models the effects of cybersecurity strategies on business performance.

Compartmental models, commonly used in epidemiology to describe disease spread, can be employed to represent entity transitions between different discrete states over time. Mathematical equations define the transitions, which may depend on the size of other compartments and on static transition parameters. In \cite{roscoe2020simulation}, this approach is applied to malware propagation in Connected and Autonomous Vehicles. The SIR model, where entities are either susceptible to infection, infected, or recovered from infection, is used to examine cases where infected vehicles can be permanently patched, preventing reinfection. On the other hand, the SIS model represents the scenarios where vehicles, once recovered, are susceptible to reinfection. Similarly, in \cite{kharabsheh2024seir} the SEIR model is used to include also the condition where a device is \textit{exposed}, enabling the capture of the latency period of infections, as threats often remain undetected before causing harm.

\subsection{Petri Nets}
Petri nets are directed bipartite graphs with two types of elements: \textit{places}, representing states or system resources, and \textit{transitions}, which are events or actions triggered by the accomplishment of certain conditions. Furthermore, places may contain \textit{tokens}, which represent any type of resources, and \textit{arcs} defining the flow between places and transitions.
Despite being designed to model distributed systems, this approach can also be applied in cybersecurity, allowing the representation of the sequence of concurrent actions that occur during a cyberattack.  
In \cite{mayfield2019component}, Petri nets are used to represent attacker strategies, system vulnerabilities, and defensive mechanisms. Attack patterns are drawn from the Common Attack Pattern Enumeration and Classification database and formalized as modular Petri net components, allowing for the representation of adversarial actions such as privilege escalation and service disruption. The tool supports both fine-grained modeling, where components represent specific attack techniques within different phases (e.g., exploration, exploitation), and coarse-grained modeling, where components encapsulate the entire attack patterns. 

Petri Nets could also be extended to support more expressive modeling capabilities, such as incorporating time, data, costs, and strategic behavior.
In \cite{tritilanunt2006using}, Timed Coloured Petri Nets, where tokens carry data values and transitions are associated with time delays, are used to assess the resilience of protocols like SSL and HIP against \ac{DoS}. A cost-based model quantifies the computational burden on different protocol participants, providing insights into the effectiveness of mitigation strategies. The simulation analyzes execution costs, attack impact, and overall system robustness. 
In \cite{petty2022modeling}, the authors introduce \acp{PNPSC}, where competing players act based on the part of the system they can observe, taking into account the cost required to modify or perform those actions. The methodology is applied to generate executable models of real-world attack patterns from the CAPEC database, enabling reinforcement learning agents to learn cost-efficient strategies based on the observed outcomes.

\subsection{Hardaware-based Simulation}
Usually, simulations rely on a certain level of abstraction, neglecting the physical features of devices and communication infrastructures. However, certain scenarios may require the integration of real hardware into the testing environment, in order to observe the real behavior of devices under attack and to evaluate hardware-specific vulnerabilities, which are often hard to model. 

\ac{HIL} is an approach that links physical devices, such as routers or sensors, with software models that represent the rest of the investigated environment. This methodology is particularly valuable in fields where real-time performance and hardware-dependent behaviors are particularly critical, such as automotive systems~\cite{lee2023simulation} and \acp{CPS}~\cite{potteiger2017evaluating}. 
Also, \acp{WAMS} monitor are an attractive target for cyber-attacks, as they monitor and control modern power grids. In \cite{adhikari2016wams}, \ac{HIL} is used to integrate physical devices such as relays and phasor measurement units into a power system model that analyzes cyber threats such as \ac{MitM}, replay attacks, and data injection against \ac{WAMS}. A similar approach is adopted in~\cite{hou2024simulation} to simulate data injection attacks against the secondary control of DC microgrids, and to study their impact on voltage stability and current control. Finally, in \cite{sun2022hil} \ac{HIL} is used to perform tests against \ac{DDoS}, considering different network topologies.

While \ac{HIL} relies on physical components to test embedded systems, the native simulation runs real embedded software on a virtualized hardware model.
This approach eliminates the need to employ actual devices, allowing for faster, scalable, and flexible testing while maintaining accuracy in power consumption, execution time, and network behavior estimations. In \cite{diaz2016simulation}, native simulation is used to analyze the effects of attacks on \acp{WSN}, considering components like hardware (e.g.,~processors, sensors, and RF transceivers), embedded software, and network deployment. 
Finally, this methodology is used in \cite{rothbart2004high} to represent fault injection techniques on smart cards at the functional level. Faults are introduced into interconnections and memory by modifying data exchanges between components, allowing the assessment of vulnerabilities without altering the hardware.

\subsection{Probabilistic Modeling}
To capture uncertainty and variability in real-world systems, modelers can leverage mathematical frameworks that use probability distributions to represent stochastic behavior and estimate the likelihood of possible outcomes.

Monte Carlo simulation is a technique for analyzing complex systems affected by uncertainty. It generates random failure and repair events based on statistical distributions and provides a probabilistic evaluation of system reliability and performance over time.
In, \cite{tatar2016impact}, Monte Carlo simulations are performed to quantify the impact of \ac{DoS} and attacks targeting the integrity of controls and sensor signals on power generation systems, considering the financial losses. By generating random failure and repair times based on probabilistic distributions, modelers can assess how attacks affect system availability and downtime.
Bayesian methods represent another probabilistic approach, integrating prior knowledge into the analysis and updating probability estimates as new information is obtained. 
Bayesian attack trees model cyber-physical security risks by analyzing attack paths, dependencies, and intrusion probabilities based on attacker skill and exploit difficulty.
In \cite{meyur2020bayesian}, this method is used to evaluate vulnerabilities in power systems, focusing on threats like \ac{DoS} and \ac{XSS}.
The study analyzes how attacks on SCADA systems in substations and control centers compromise Human Machine Interfaces, application servers, and other network components. The Bayesian model quantifies exploit probabilities based on vulnerability type, complexity, and attacker skill level, providing an assessment of security risks.
In \cite{futuransky2009}, the likelihood of exploit success is determined by probability distributions based on system attributes like OS version and security configurations rather than simulating the detailed execution of exploits. Specifically, the proposed simulation platform enables the simulation of vulnerabilities, exploits, and complex multi-step attacks (i.e. use of a compromised machine as a stepping stone to reach further networks and machines, making use of its trust relationships).

\subsection{Other Modeling Approaches}
While most of the investigations employed \ac{ABM} or the other mentioned modeling paradigms, other approaches can be found in the state of the art.

Dynamic node-level modeling is a methodology that is used to analyze the spread of cyber threats by representing each node in a network as an individual entity with state-based transitions. Unlike traditional epidemic models that assume uniform spreading behavior, this approach accounts for network topology via an adjacency matrix, enabling a more granular analysis of infection dynamics.
In \cite{liu2019modeling}, a dynamic node-level model is used to study ransomware propagation, where node transitions between susceptible, delitescent, infected, and recovered states are based on probabilistic rules influenced by neighboring nodes. Theoretical analysis leverages spectral properties of the adjacency matrix to determine conditions for controlling ransomware spread. Numerical simulations on different network topologies, including star, fully connected, and scale-free networks, show the impact of network structure on infection rates and containment strategies.

G-networks employ a queueing-based modeling approach for analyzing complex networked systems under dynamic conditions, using positive arrivals to represent normal data flows and negative arrivals to model disruptive events such as cyberattacks. 
In \cite{sarigiannidis2017modeling}, a G-network is used to assess security threats in \ac{IoT} infrastructures, estimating data losses in relation to traffic arrival rates. The study highlights the role of queuing dynamics in IoT resilience, showing how variations in arrival and departure rates influence data availability in the application domain.

Reinforcement learning is mainly used to optimize decision-making, but it can be applied in cybersecurity simulations to model attack scenarios where agents interact with the system, receive feedback, and gradually learns effective strategies. This method is useful for penetration testing, allowing agents to explore different ways to exploit vulnerabilities without predefined rules. 
In \cite{erdHodi2021simulating}, SQL injection is modeled as a learning challenge, where agents interact with a vulnerable system, observe responses to their inputs, and learn how to craft effective SQL injection queries to extract sensitive data. Two reinforcement learning approaches are compared: \textit{tabular Q-learning}, where the agent’s behavior is easier to interpret but struggles with scalability, and \textit{deep Q-learning}, which uses neural networks to handle larger action spaces more effectively. 

\subsection{Multilevel Modeling}
One model, regardless of the employed approach, is often incapable of capturing all the relevant aspects of complex scenarios. Real-world systems are frequently composed of multiple interacting components, each addressing semantically distinct aspects that contribute to overall system behavior.
An example can be found in \cite{murillo2020co}, where a hydraulic simulator and an industrial network emulator are combined to reproduce \ac{MitM} attacks in water distribution systems.
Multilevel modeling is widely used to integrate the various building blocks, particularly when it is possible to reuse existing simulators, whose effectiveness in representing specific phenomena has already been validated. A significant example is SURE~\cite{koutsoukos2017sure}, a multilevel simulator designed to assess resilience and security of \acp{CPS}, with a focus on smart transportation. The tool combines OMNeT++ to simulate communication networks, SUMO to model traffic dynamics, and Matlab/Simulink to run control algorithms that govern the transportation infrastructure. 
A similar approach was applied in~\cite{cheng2022modeling} to analyze the impact of cyberattacks on heterogeneous intelligent traffic flow. A car-following model representing microscopic driving dynamics is integrated with a communication model that captures information exchange in connected vehicle systems. In particular, the investigation examines bogus messages, replay/delay attacks, and collusion attacks, which all use tampered information to disrupt cooperative driving, causing instability in traffic flow.

In multilevel modeling, different modeling paradigms may be combined to represent processes that evolve with different rules, integrating models that use continuous time and space with discrete representations. For instance, in \cite{fu2021modeling}, continuous-time models represent the heating, ventilation, and air conditioning physical processes, while discrete-event models capture the building automation system communication network, allowing the simulation of data intrusion and denial-of-service attacks targeting the communication infrastructure and their effects on system performance and stability.
Continuous and discrete representations are also combined in \cite{ciancamerla2013modeling}, to analyze the impact of cyberattacks on a medium-voltage electrical grid. NetLogo is used to simulate malware propagation across the corporate network and SCADA devices using a SIR model, while NS2 models \ac{DoS} and \ac{MITM} attacks on SCADA communications. The study assesses how these attacks lead to Loss of View and Loss of Control, disrupting communication between the SCADA control center and Remote Terminal Units.

In \cite{yao2024simulation}, a modular approach is employed to simulate side channel attacks.
The simulator is composed of multiple building blocks: \textit{data acquisition} module to capture side channel information from the target device, \textit{data processing} module to process the acquired data, \textit{leakage} module to generate synthetic side-channel signals, and \textit{attack strategy} module to implement the attacks. Similarly, in \cite{liu2018real} a modular approach is employed to represent \ac{MitM} and \ac{DoS} attacks in a smart grid, representing aspects such as power system, communication system, and decision making. The integration of \ac{HIL} enables a real-time testing of how actual control and protection devices respond to cyberattacks.

Despite the benefits of combining multiple existing simulators, modelers must be very careful to appropriately couple the various atomic components, synchronizing execution and ensuring consistent data transmission.
In \cite{top2017simulation}, the ParGrid platform orchestrates the real-time interaction between a power system simulator (GridDyn) and a network simulator (ns-3), enabling the simultaneous analysis of power grid behavior and SCADA communications. Cyberattacks are introduced at the communication layer, affecting Remote Terminal Unit operations and propagating their impact on the power system, where consequences such as breaker misoperations or grid instability are analyzed. 
Similarly, in~\cite{de2020co}, the Mosaik framework orchestrates the execution of ns-3 and a power system simulator (OpenDSS), in order to analyze the impact of cyberattacks on smart grid operations, synchronizing data exchange and time-steps between the simulators. 

Finally, simulation and emulation can be combined to exploit their respective strengths, balancing scalability with realism. In \cite{vaughn2016addressing}, this approach is applied to create a high-fidelity testbed for Industrial Control Systems. Emulation incorporates real devices and protocols to ensure accurate network interactions, while simulation improves scalability by modeling SCADA operations, physical processes, and communication networks.

\subsection{Discussion}

\begin{table}[ht]
\centering
\caption{Modeling techniques and key characteristics}
\small
\begin{tabular}{|p{0.28\textwidth}|p{0.18\textwidth}|p{0.18\textwidth}|p{0.18\textwidth}|}
\hline
\textbf{Technique} & \textbf{Detail Level} & \textbf{Stochastic} & \textbf{Dynamic} \\
\hline
Agent-Based Modeling & High & Possible & Yes \\
\hline
System Dynamics & Low & Yes & No \\
\hline
Petri Nets & Medium & No & Yes \\
\hline
Hardware Simulation & Very High & No & Yes \\
\hline
Probabilistic Modeling & Low & Yes & No \\
\hline
\end{tabular}
\label{tab:modeling-techniques}
\end{table}

As summarized in Table~\ref{tab:modeling-techniques}, the choice of a simulation approach depends on several factors, including the required level of accuracy, data utilization, the complexity of the tested environment, and the incorporation of stochastic and dynamic elements.
The accuracy of the representation usually depends on the objectives of the simulation and computational constraints. Fine-grained models reproduce the behavior of all the entities involved, considering their heterogeneity. Low-level characteristics can be taken into account, such as network protocols, software vulnerabilities, security controls, and human-related decisions, allowing for testing how specific configurations of the target systems or attack features influence the outcome of cyber intrusion.
On the other hand, coarse-grained models focus on attack dynamics at the macro scale, such as malware spreading across networks or coordinated intrusion attempts, prioritizing efficiency over accuracy.
In certain models, actors operate according to predefined rules, where attack actions and system responses are explicitly defined, making these models useful for evaluating known vulnerabilities and testing specific defense mechanisms. On the other hand, other approaches incorporate stochasticity, introducing variations to better reproduce the unpredictability of real-world systems.
Another important factor is whether the simulation accounts for adaptive behavior. Some models assume fixed attack strategies, while others allow attackers and defenders to adjust their behavior based on accumulated knowledge. This is critical when analyzing threats that evolve over time, such as adversaries modifying their attack patterns to evade detection or malware that changes its characteristics to bypass security measures.
To increase the veracity of the simulation, modelers could leverage data extracted from historical intrusion records or synthesized using statistical models to introduce realistic variations. 

From a technological point of view, certain simulators are frequently employed for cybersecurity research, particularly ns-3 and OMNeT++ for modeling communication over networks, Mininet for representing \ac{SDN}, and SUMO for traffic simulation. Other times, authors develop novel simulators from scratch, though only rarely do they provide the source code. In terms of programming languages, the most frequently employed are C, C++, Java, Python, and Matlab.

\section{Goal of the Simulation}\label{sec:goals}
The ultimate goal of the simulation can influence the design choices of researchers, including the modeling paradigms employed, the adopted level of granularity, the system configurations to be tested, and the metrics used for evaluation.
In this section, we discuss the interconnection between the goal of the investigation and the simulation approach.

\begin{figure}[h]
\centering\includegraphics[scale=\defaultscale]{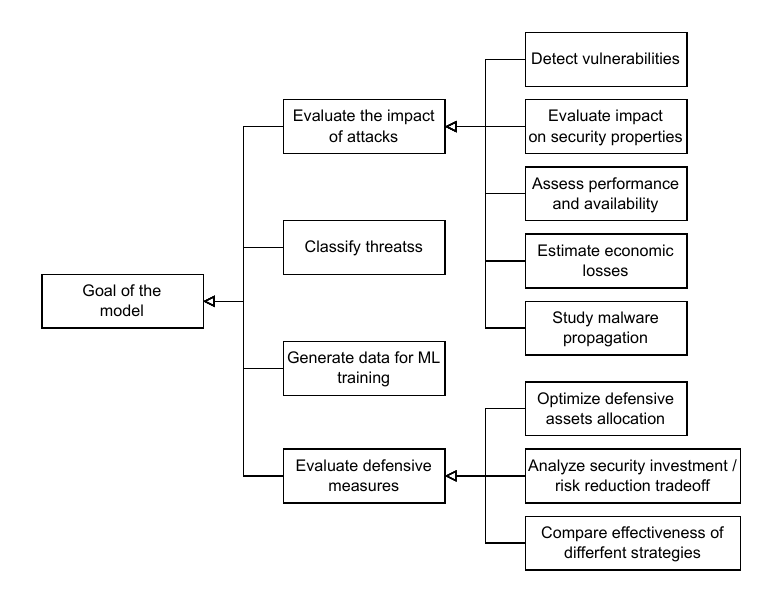}
\caption{Taxonomy of the goals of the simulation.}
\label{taxonomy-goal}
\end{figure}

\subsection{Evaluate Attack Impact and System Resilience}
A primary objective of cybersecurity simulations is to assess the system's exposure to cyberattacks, analyzing its performance under adversarial conditions and the capacity to recover from failures. This includes the evaluation of system's vulnerabilities to known cyber threats, measuring the degree of operational degradation under adversarial conditions, examining the potential cascading effects of disruption and how malware infection can propagate across interconnected components. 
In~\cite{carraminana2023enhancing}, the authors analyze how disruptions to information and communication systems affect hospital operations. The simulation evaluates the extent of operational degradation, focusing on hospital service availability, patient admissions, and resource utilization under different attack scenarios, including \ac{DoS} and ransomware, ultimately assessing how these factors contribute to increased mortality rates.

Simulations also help identify critical thresholds below which the system functionalities are significantly compromised, and what conditions lead to irreversible failures or prolonged downtime.
For instance, in~\cite{serena2022security}, the outcome of attacks on Proof-of-Work-based blockchains is evaluated under varying attack conditions, identifying critical thresholds for system security and operation. Specifically, core parameters are the computational power controlled by the attackers in the context of 51\% attacks and selfish mining, and the number of malicious nodes for the Sybil attack (see Section~\ref{sec:blockchain}).

The outcome of cyberattacks is not influenced only by adversarial features, but also by multiple environmental factors such as defensive policies or network organization. As a result, simulation can take into account various system layouts to understand which system configurations increase the likelihood of a successful attack.
In~\cite{long2005denial}, the authors have studied how packet loss and delay jitter are related with performance degradation of network-based control systems under \ac{DoS} attacks. Two queuing models are used to simulate different attack scenarios: \textit{local DoS} targeting endpoint devices, leading to packet losses, and \textit{non-local DoS} affecting intermediate routers, causing prolonged delay jitter. Results are expressed with metrics such as percentage overshoot, rise and settling time, and mean-squared error, showing that defensive measures at the network level can mitigate performance degradation.
In~\cite{kouril2014cloud}, the authors propose a cloud-based testbed to evaluate the impact on networked infrastructures. Through virtualization, it enables researchers to represent threats such as \ac{DoS}, and analyze system performance under different network configurations. 
\ac{ABM} is used in~\cite{bayir2020company} to analyze which are the most impactful factors on the security of mid-size and small companies. 
The developed platform allows users to set parameters such as network size, attack strength and frequency, number and strength of defensive agents, and number of end devices, in order to assess the company’s ability to maintain operational stability during and after an attack.
A broader socio-technical perspective is adopted in~\cite{charitoudi2014agent} to evaluate the security impact on complex organizations' networks. In particular, the model incorporates the roles of the human agents, defined as the sets of rights and responsibilities, expectations, behaviors, or expected behaviors and norms. The simulation reproduces the consequences that occur when a compromised agent cannot fulfill a responsibility, analyzing how cascading effects propagate through business processes, technical systems, and human decision-making.

Certain types of malware, such as worms and viruses, do not aim to infect a single device or system, but they attempt to spread across multiple nodes, exploiting network connectivity and user interactions. The speed and capillarity of dissemination depend on several factors, including user behavior, security policies, and the structural properties of the network. Actions such as opening malicious files or forwarding infected messages strongly boost the propagation, while network topology and access control policies influence how malware moves between nodes.
In~\cite{garetto2003modeling}, the propagation of email-borne viruses is investigated using a stochastic model based on Interactive Markov Chains. The study combines analytical techniques and discrete event simulations to assess the speed and extent of malware dissemination, considering factors like user behavior (e.g.,~opening infected emails) and network topology.

Finally, economic factors play a crucial role in ransomware attacks, influencing both attacker strategies and the decisions of the targeted system. Models must account not only for the technical impact of ransomware, but also for the financial consequences and decision-making policies that organizations can adopt.
In~\cite{skeoch2022modelling}, a Partially Observable Markov Decision Process is used to simulate ransomware attacks and evaluate economic decision-making in response to infections. The model considers uncertainty in system states, allowing organizations to weigh different actions, such as paying the ransom, attempting to repair the system, or shutting down operations. 

\subsection{Evaluate Countermeasures}
Multiple countermeasures can be implemented to mitigate the effect of cyberattacks, and simulation can be employed to compare their effectiveness, helping identify the most suitable solution in a certain context.
While certain defensive strategies are thought to mitigate specific threats, others are designed to protect the targeted system as a whole. 
Typical defensive mechanisms include \acp{WAF}, \acp{IPS}, \acp{IDS}, and \acp{EDR}~\cite{roque2024implementation}. These solutions vary in scope and functionality: \acp{WAF} protect web applications by monitoring and filtering HTTP traffic, \acp{IDS} analyze network traffic to detect malicious activity without direct intervention, \acp{IPS} actively block detected threats, and \acp{EDR} focus on endpoint security by detecting, investigating, and responding to threats at the device level.
In~\cite{sanodiya2017attacks}, multiple defensive mechanisms against~\ac{DoS} have been considered, such as Pushback (i.e.,~when congestion is detected, routers limit the flow by blocking suspicious packets upstream), \ac{SIFF} (i.e.,~a network mechanism that establishes reserved communication channels, ensures packet authenticity, and filters traffic independently of flow states), \ac{AITF} (i.e.,~a victim requests its gateway to block malicious traffic and collaborates with the attacker’s nearest router to enforce packet filtering), and \ac{TVA} (i.e.,~network traffic is regulated through a token-based validation).
In~\cite{wagner2016towards}, an automated decision-support system assists in the selection of effective network segmentation strategies. Network segmentation is a security measure that splits a network into isolated fragments, regulating communication between the various parts to limit unauthorized access and contain the propagation of security breaches. 
In~\cite{albarakati2022multi}, both natural and cyber-induced faults are introduced into a power distribution system to observe how the system detects, locates, and isolates them, and eventually recovers. Agents monitor current measurements and breaker statuses, exchanging data to identify fault locations and determine the nature of disruptions. A key metric is fault localization time, which measures how quickly the system identifies the location of faults or attacks.

Cybersecurity does not depend only on technological features but also on human factors, as cyber situation awareness (i.e.,~the ability of defenders to perceive, understand, and anticipate threats) is crucial for detecting the attacks that exploit cognitive limitations, such as delayed anomaly detection or misjudgment of risk.
In~\cite{dutt2013cyber}, a simulator based on Instance-Based Learning Theory (i.e.,~a theory that explains how people make decisions based on past experiences, using stored examples to handle dynamic and complex situations) is used to assess defenders' awareness against impatient attacks, where threats appear early to overwhelm them, and patient attacks, where threats are delayed to evade detection. The simulation processes sequences of network events through memory-based mechanisms, where decisions are influenced by the similarity, frequency, and recency of past experiences.

Besides assessing the effectiveness of defensive mechanisms against cyber threats, M\&S may also focus on the associated costs, including deployment, maintenance, and resource consumption.
The cost-effectiveness trade-off in Industrial Control Systems is evaluated in~\cite{fielder2016modelling}, analyzing the balance between security investment and risk reduction.
The model reflects how early investments in defense provide significant protection gains, while additional spending yields progressively smaller improvements. The diminishing return effect highlights the importance of strategic allocation during the design phase.

Implementing security measures might increase the protection of a system against attacks, but defensive assets must be positioned strategically in order to minimize unnecessary costs or overhead.
Simulation allows researchers to test various configurations, helping identify the best policy to balance protection, cost, and operational efficiency.
In~\cite{grabis2019simulation}, \ac{ABM} is used to evaluate and optimize the allocation of officers in a community-based fraud detection system designed to identify fraudulent internet resources. The system follows a three-tier structure, where Level-1 officers, drawn from an open pool of internet users, issue initial verdicts on reported fraud cases. Their evaluations are then reviewed by Level-2 officers, who assess and validate the decisions, while Level-3 officers make the final determination based on accumulated approvals.

\subsection{Assist ML models}
Other than evaluating how a system responds to cyber attacks and the effectiveness of defense measures, simulation can be combined with machine learning to conduct data-driven cybersecurity investigations.

One key application is the generation of realistic data (e.g.,~ service logs, network traffic traces, system metrics) for training machine learning models, assisting in the classification of suspicious activity~\cite{prasad2024simulation}. In fact, sometimes it is hard to obtain such data from real-world systems due to privacy concerns, operational risks, or the rarity of certain events.
This approach is frequently used to detect fraudulent activities in smart city environments. In~\cite{aljamal2024optimizing}, both legitimate clients and malicious users performing \ac{DoS} attacks are modeled. The generated network traffic data is then fed to train machine learning classifiers, which analyze patterns to distinguish between regular and malicious activity. This methodology is also applied in~\cite{de2024ddoshield} to train intrusion detection models for botnet-driven DDoS in IoT networks. Multiple machine learning models are considered, such as K-Means, Random Forest, and Convolutional Neural Networks. Similarly, in~\cite{jia2020flowguard}, simulation is used to generate traffic data for training machine learning models in both detecting and classifying IoT-related DDoS attacks. Once malicious behaviors have been detected, attacks are further classified into specific types using machine learning models trained on a combination of simulated and real attack datasets.
Classification capability can strengthen the effectiveness of defensive mechanisms like~\acp{IDS}, allowing them to recognize specific threats based on detected patterns. In~\cite{wrana2022od1nf1st}, simulation generates realistic data about military avionics communication to train an \ac{IDS} through reinforcement learning techniques.
This strategy can also be applied to generate a data log for \ac{UAV} missions. To enhance threat detection capability, in~\cite{rimoli2024synthetic} simulation introduces various anomalies, including sensor failures, component malfunctions, and communication attacks.

In cybersecurity, AI not only refers to classifiers that identify suspicious patterns, but machine learning techniques can be used to make strategic decisions in an adversarial scenario.
In~\cite{carrasco2024cybershield}, data generated from the simulation is used to support the training of Reinforcement Learning algorithms for cybersecurity. In the architecture, both the main network and a honeynet are represented, with hosts having customizable features, including vulnerabilities drawn from the National Vulnerability Database. 

Simulation can be used to recreate realistic attack scenarios and system behaviors, providing a controlled environment where machine learning techniques can be applied to detect patterns, make predictions, and evaluate security measures.
In~\cite{ashiku2020agent}, \ac{ABM} is integrated with a neural network to assess cybersecurity risk for business entities. The simulation represents network components, attack propagation, and defense mechanisms, while the neural network, trained on the UNSW-NB15 dataset, classifies attacks based on network traffic features. The output of the ML model then updates asset states in the simulation, influencing attack spread and defense activation. A predator-prey model engages the best-performing defense mechanisms based on detected threats, allowing dynamic risk assessment and cost evaluation.

Finally, neural networks can be integrated into a digital twin platform to enable real-time process control with the aim to enhance the safety
and reliability of the physical plant. For instance, in~\cite{ali2024development} an \ac{MPC}-based digital twin is developed for a fused deposition modeling printer, integrating machine learning-driven anomaly detection to safeguard against cyber threats. The neural network processes sensor data in real time to identify anomalies in the printing process, such as deviations in nozzle temperature, bed temperature, extruder vibration, and position shift.

\subsection{Others}
While the discussed objectives are recurrent in cybersecurity studies, certain investigations do not fall within the mentioned categories.
For instance, simulation can be used to evaluate the effectiveness and accuracy of testing tools or experimental setups.
In~\cite{malekzadeh2011validating}, the reliability of OMNeT++ in simulating \ac{DoS} attacks on wireless networks was evaluated by comparing simulation results with a real testbed. The study focused on attacks targeting control frames (ACK, RTS, CTS), assessing their impact on network performance through key metrics such as end-to-end delay, throughput, and packet loss. The OMNeT++ extension module developed for this purpose was validated against experimental results, showing a high degree of accuracy under heavy load conditions.
Simulation can also be used to assess the effectiveness of specific metrics in detecting cyberattacks. By modeling different attack scenarios and analyzing system responses, simulations can help determine whether certain parameters serve as reliable indicators of malicious activity. In~\cite{eastman2017simulation}, packet loss and energy consumption are evaluated as potential metrics for intrusion detection in IoT networks. 

Simulation can also support classification tasks, distinguishing between different types of attacks based on the observable behavior.
In~\cite{hussain2003framework}, synthetic attack streams are generated based on attack dynamics observed in real-world traffic to validate a classifier that distinguishes between single-source \ac{DoS} and \ac{DDoS}. The classifier first analyzes packet headers, then checks how the attack ramps up over time, and finally analyzes the traffic’s spectral content. Simulations confirm that multi-source attacks, due to a lack of synchronization, exhibit lower-frequency spectral concentration. 
Similarly, in~\cite{zhang2021network}, the purpose of the simulation is to recognize the real intention behind network attacks. Interactions between attackers and defenders are modeled as a signaling game, where the defender interprets attack behaviors to infer the attacker’s true objective.

Finally, simulation can help evaluate cyber risks in real-time, flagging suspicious activities and identifying and classifying cyberattacks like \ac{DDoS} in satellite communication~\cite{mashayekhi2024simulation}.

\subsection{Discussion}

\begin{table}[ht]
\centering
\caption{Key characteristics depending on the goal of the simulation}
\small
\begin{tabular}{|p{0.28\textwidth}|p{0.28\textwidth}|p{0.28\textwidth}|}
\hline
\textbf{Goal} & \textbf{Outputs or metrics} & \textbf{Modeling Focus} \\
\hline
\makecell{Attack impact \\ \& system resilience} & \makecell{Latency \\ Packet loss \\ Service downtime} & \makecell{Predict effects of attacks \\ Discover vulnerabilities \\ Threshold analysis} \\
\hline
\makecell{Defensive \\ measures evaluation} & \makecell{Detection rate \\ Delay mitigation \\ Resource overhead} & \makecell{Compare strategies \\ Assess countermeasures} \\
\hline
\makecell{Attack \\ propagation} & \makecell{Infection rate \\ Peak spread time} & \makecell{Model infection \\ mechanisms} \\
\hline
\makecell{ML \\ training} & \makecell{Synthetic logs} & \makecell{Generate realistic \\ labeled data} \\
\hline
\makecell{Economic \\ analysis} & \makecell{Expected \\ economic losses} & \makecell{Correlate attack intensity \\ with economic impact} \\
\hline
\end{tabular}
\label{tab:goals}
\end{table}

The goal of the simulation drives the model design, determining its focus as well as the metrics and parameters to be used, as shown in Table~\ref{tab:goals}.
In most cases, simulations have two main objectives: evaluating the resilience of a system under attack conditions and assessing the effectiveness of defensive mechanisms. 
These goals are often interdependent, as the ability to represent the consequences of attacks provides the foundation to evaluate the effectiveness of security measures in response.
When examining the impact of cyberattacks, simulations seek to quantify the extent of system degradation under different conditions. By varying attack parameters and system configurations, simulations help researchers assess worst-case scenarios and identify critical failure points. 
Depending on the context, different metrics may describe the level of impairment of a system, such as response time, resource consumption, service availability, data integrity, system recovery time, and economic losses. 
Analogously, these metrics can be employed to evaluate and compare different mitigation strategies, where an additional important factor is the ability and the time to detect malicious patterns.
Given the close relationship between attack impact and defense performance, many simulations integrate both aspects into a single framework. An attack scenario can serve not only to measure system resilience but also to test how various countermeasures perform in reducing damage and restoring normal operations.
Finally, the ability to customize system parameters makes simulation particularly suitable to generate data that are used to train machine learning models, which typically require a large volume of labeled data that possibly covers a wide range of attack scenarios and operating conditions.

\section{Conclusions}\label{sec:conclusions}
Simulation is a widely employed approach for cybersecurity investigations, providing a controlled environment to analyze threats, test defense mechanisms, and assess system resilience by evaluating the impact of cyberattacks on network stability, resource consumption, operational availability, data integrity, and confidentiality.
Over the last $25$ years, researchers have carried out several studies, which differ for the considered application domain, the cyber threats analyzed, the modeling techniques employed, and the goal of the investigations.

However, despite its benefits, simulation-based research has limitations. So\-me models fail to fully capture the complexity of cyberattacks, while human factors, such as user behavior and employee decision-making, remain challenging to represent with accuracy.
Thus, it turns out that simulation is more suited for modeling attacks that have a measurable, system-level impact, such as \ac{DoS}, where variables like traffic volume, resource consumption, and service availability can be easily quantified.


\begin{thebibliography}{100}

\bibitem{burger2019estimating}
Olga B{\"u}rger, Bj{\"o}rn H{\"a}ckel, Philip Karnebogen, and Jannick
  T{\"o}ppel.
\newblock Estimating the impact of it security incidents in digitized
  production environments.
\newblock {\em Decision Support Systems}, 127:113144, 2019.

\bibitem{cetinkaya2019overview}
Ahmet Cetinkaya, Hideaki Ishii, and Tomohisa Hayakawa.
\newblock An overview on denial-of-service attacks in control systems: Attack
  models and security analyses.
\newblock {\em Entropy}, 21(2):210, 2019.

\bibitem{conti2016survey}
Mauro Conti, Nicola Dragoni, and Viktor Lesyk.
\newblock A survey of man in the middle attacks.
\newblock {\em IEEE communications surveys \& tutorials}, 18(3):2027--2051,
  2016.

\bibitem{yohanandhan2020cyber}
Rajaa~Vikhram Yohanandhan, Rajvikram~Madurai Elavarasan, Premkumar Manoharan,
  and Lucian Mihet-Popa.
\newblock Cyber-physical power system (cpps): A review on modeling, simulation,
  and analysis with cyber security applications.
\newblock {\em IEEE Access}, 8:151019--151064, 2020.

\bibitem{vestad2024survey}
Arnstein Vestad and Bian Yang.
\newblock A survey of agent-based modeling for cybersecurity.
\newblock {\em Human Factors in Cybersecurity}, 127(127), 2024.

\bibitem{jawad2021modeling}
Alvi Jawad and Jason Jaskolka.
\newblock Modeling and simulation approaches for cybersecurity impact analysis:
  State-of-the-art.
\newblock In {\em 2021 Annual Modeling and Simulation Conference (ANNSIM)},
  pages 1--12, Los Alamitos, CA, USA, 2021. IEEE.

\bibitem{chowdhury2021cyber}
Nabin Chowdhury and Vasileios Gkioulos.
\newblock Cyber security training for critical infrastructure protection: A
  literature review.
\newblock {\em Computer Science Review}, 40:100361, 2021.

\bibitem{kavak2021simulation}
Hamdi Kavak, Jose~J Padilla, Daniele Vernon-Bido, Saikou~Y Diallo, Ross Gore,
  and Sachin Shetty.
\newblock Simulation for cybersecurity: state of the art and future directions.
\newblock {\em Journal of Cybersecurity}, 7(1):tyab005, 2021.

\bibitem{engstrom2022two}
Viktor Engstr{\"o}m and Robert Lagerstr{\"o}m.
\newblock Two decades of cyberattack simulations: A systematic literature
  review.
\newblock {\em Computers \& Security}, 116:102681, 2022.

\bibitem{veksler2018simulations}
Vladislav~D Veksler, Norbou Buchler, Blaine~E Hoffman, Daniel~N Cassenti, Char
  Sample, and Shridat Sugrim.
\newblock Simulations in cyber-security: a review of cognitive modeling of
  network attackers, defenders, and users.
\newblock {\em Frontiers in psychology}, 9:691, 2018.

\bibitem{keele2007guidelines}
Staffs Keele et~al.
\newblock Guidelines for performing systematic literature reviews in software
  engineering.
\newblock Technical report, Technical report, ver. 2.3 ebse technical report.
  ebse, 2007.

\bibitem{zhao2022design}
Yongliang Zhao, Rongheng Lin, Hua Zou, and Wei Zeng.
\newblock Design and implementation of ddos attack and defense simulation
  subsystem based on mininet.
\newblock In {\em 2022 IEEE 22nd International Conference on Communication
  Technology (ICCT)}, pages 1328--1334, Los Alamitos, CA, USA, 2022. IEEE.

\bibitem{kotenko2006simulation}
Igor Kotenko and Alexander Ulanov.
\newblock Simulation of internet ddos attacks and defense.
\newblock In {\em Proceedings of the 9th International Conference on
  Information Security}, ISC'06, page 327–342, Berlin, Heidelberg, 2006.
  Springer-Verlag.

\bibitem{novak2017modeling}
Christopher Novak, Jim Blythe, Ross Koppel, Vijay~H. Kothari, and Sean~W.
  Smith.
\newblock Modeling aggregate security with user agents that employ password
  memorization techniques.
\newblock In {\em Symposium On Usable Privacy and Security}, Santa Clara, CA,
  2017. USENIX Association.

\bibitem{karagiannis2023cybersecurity}
Stylianos Karagiannis, Emmanouil Magkos, Christoforos Ntantogian, Ricardo
  Cabecinha, and Theofanis Fotis.
\newblock Cybersecurity and medical imaging: A simulation-based approach to
  dicom communication.
\newblock {\em Applied Sciences}, 13(18):10072, 2023.

\bibitem{thompson2018agent}
Brian Thompson and James Morris-King.
\newblock An agent-based modeling framework for cybersecurity in mobile
  tactical networks.
\newblock {\em The Journal of Defense Modeling and Simulation}, 15(2):205--218,
  2018.

\bibitem{dobson2017cyber}
Geoffrey~B. Dobson and Kathleen~M. Carley.
\newblock Cyber-fit: An agent-based modelling approach to simulating cyber
  warfare.
\newblock In Dongwon Lee, Yu-Ru Lin, Nathaniel Osgood, and Robert Thomson,
  editors, {\em Social, Cultural, and Behavioral Modeling}, pages 139--148,
  Cham, 2017. Springer International Publishing.

\bibitem{daah2025simulation}
Clement Daah, Amna Qureshi, Irfan Awan, and Savas Konur.
\newblock Simulation-based evaluation of advanced threat detection and response
  in financial industry networks using zero trust and blockchain technology.
\newblock {\em Simulation Modelling Practice and Theory}, 138:103027, 2025.

\bibitem{moskal2018cyber}
Stephen Moskal, Shanchieh~Jay Yang, and Michael~E Kuhl.
\newblock Cyber threat assessment via attack scenario simulation using an
  integrated adversary and network modeling approach.
\newblock {\em The Journal of Defense Modeling and Simulation}, 15(1):13--29,
  2018.

\bibitem{qwasmi2011simulation}
Nidal Qwasmi, Fayyaz Ahmed, and Ramiro Liscano.
\newblock simulation of ddos attacks on p2p networks.
\newblock In {\em 2011 IEEE International Conference on High Performance
  Computing and Communications}, pages 610--614, Los Alamitos, CA, USA, 2011.
  IEEE.

\bibitem{vasenin2014environment}
Valeriy Vasenin, Vladimir Roganov, and Andrey Zenzinov.
\newblock Environment for hybrid simulation of information security solutions
  for grid and cloud-systems.
\newblock In {\em 2014 4th International Conference On Simulation And Modeling
  Methodologies, Technologies And Applications (SIMULTECH)}, pages 255--260,
  Los Alamitos, CA, USA, 2014. IEEE.

\bibitem{le2020gridattacksim}
Tan~Duy Le, Adnan Anwar, Seng~W Loke, Razvan Beuran, and Yasuo Tan.
\newblock Gridattacksim: A cyber attack simulation framework for smart grids.
\newblock {\em Electronics}, 9(8):1218, 2020.

\bibitem{chen2013impact}
Bo~Chen, Karen~L. Butler-Purry, and Deepa Kundur.
\newblock Impact analysis of transient stability due to cyber attack on facts
  devices.
\newblock In {\em 2013 North American Power Symposium (NAPS)}, pages 1--6, Los
  Alamitos, CA, USA, 2013. IEEE.

\bibitem{stefanov2014cyber}
Alexandru Stefanov and Chen-Ching Liu.
\newblock Cyber-physical system security and impact analysis.
\newblock {\em IFAC Proceedings Volumes}, 47(3):11238--11243, 2014.

\bibitem{stuanculescu2021case}
Marilena St{\u{a}}nculescu, Sorin Deleanu, Paul~Cristian Andrei, and Horia
  Andrei.
\newblock A case study of an industrial power plant under cyberattack:
  Simulation and analysis.
\newblock {\em Energies}, 14(9):2568, 2021.

\bibitem{ding2017attacks}
Peili Ding, Yinan Wang, Gangfeng Yan, and Wei Li.
\newblock Dos attacks in electrical cyber-physical systems: A case study using
  truetime simulation tool.
\newblock In {\em 2017 Chinese Automation Congress (CAC)}, pages 6392--6396,
  Los Alamitos, CA, USA, 2017. IEEE.

\bibitem{cunningham2017adapting}
Christine Cunningham and Antonio Roque.
\newblock Adapting an agent-based model of socio-technical systems to analyze
  security failures.
\newblock In {\em 2017 IEEE International Symposium on Technologies for
  Homeland Security (HST)}, pages 1--7, Los Alamitos, CA, USA, 2017. IEEE.

\bibitem{huang2018case}
Bing Huang, Mohammad Majidi, and Ross Baldick.
\newblock Case study of power system cyber attack using cascading outage
  analysis model.
\newblock In {\em 2018 IEEE Power \& Energy Society General Meeting (PESGM)},
  pages 1--5, Los Alamitos, CA, USA, 2018. IEEE.

\bibitem{malik2020analysis}
Shahida Malik and Weiqing Sun.
\newblock Analysis and simulation of cyber attacks against connected and
  autonomous vehicles.
\newblock In {\em 2020 international conference on connected and autonomous
  driving (MetroCAD)}, pages 62--70, Los Alamitos, CA, USA, 2020. IEEE.

\bibitem{santos2023towards}
Giovanni~A Santos, Jo{\~a}o Paulo~J da~Costa, and Antonio Arlis~S da~Silva.
\newblock Towards to beyond 5g virtual environment for cybersecurity testing in
  v2x systems.
\newblock In {\em 2023 Workshop on Communication Networks and Power Systems
  (WCNPS)}, pages 1--7, Los Alamitos, CA, USA, 2023. IEEE.

\bibitem{verma2013prevention}
Karan Verma, Halabi Hasbullah, and Ashok Kumar.
\newblock Prevention of dos attacks in vanet.
\newblock {\em Wireless personal communications}, 73:95--126, 2013.

\bibitem{jackson2023agent}
Elanor Jackson, Sahra~Sedigh Sarvestani, Justin King, and Ali~R Hurson.
\newblock Agent-based modeling for analysis of cyber attacks on the intelligent
  transportation system.
\newblock In {\em 2023 IEEE 26th International Conference on Intelligent
  Transportation Systems (ITSC)}, pages 4550--4555, Los Alamitos, CA, USA,
  2023. IEEE.

\bibitem{cassou2020simulation}
Jean Cassou-Mounat, Houda Labiod, and Rida Khatoun.
\newblock Simulation of cyberattacks in its-g5 systems.
\newblock In {\em Communication Technologies for Vehicles: 15th International
  Workshop, Nets4Cars/Nets4Trains/Nets4Aircraft 2020, Bordeaux, France,
  November 16–17, 2020, Proceedings}, page 3–14, Berlin, Heidelberg, 2020.
  Springer-Verlag.

\bibitem{javaid2013uavsim}
Ahmad~Y Javaid, Weiqing Sun, and Mansoor Alam.
\newblock Uavsim: A simulation testbed for unmanned aerial vehicle network
  cyber security analysis.
\newblock In {\em 2013 ieee globecom workshops (gc wkshps)}, pages 1432--1436,
  Los Alamitos, CA, USA, 2013. IEEE.

\bibitem{puchaty2011performance}
Ethan~M Puchaty and Daniel~A DeLaurentis.
\newblock A performance study of uav-based sensor networks under cyber attack.
\newblock In {\em 2011 6th International Conference on System of Systems
  Engineering}, pages 214--219, Los Alamitos, CA, USA, 2011. IEEE.

\bibitem{ye2018analysis}
Congcong Ye, Guoqiang Li, Hongming Cai, Yonggen Gu, and Akira Fukuda.
\newblock Analysis of security in blockchain: Case study in 51\%-attack
  detecting.
\newblock In {\em 2018 5th International conference on dependable systems and
  their applications (DSA)}, pages 15--24, Los Alamitos, CA, USA, 2018. IEEE.

\bibitem{tomar2024blockchain}
Ashish Tomar.
\newblock A blockchain-based scheme to prevent the sybil attack in vanet.
\newblock In {\em 2024 15th International Conference on Computing Communication
  and Networking Technologies (ICCCNT)}, pages 1--6, Los Alamitos, CA, USA,
  2024. IEEE.

\bibitem{rajan2017sybil}
Anjana Rajan, J~Jithish, and Sriram Sankaran.
\newblock Sybil attack in iot: Modelling and defenses.
\newblock In {\em 2017 International conference on advances in computing,
  communications and informatics (ICACCI)}, pages 2323--2327, Los Alamitos, CA,
  USA, 2017. IEEE.

\bibitem{chicarino2020detection}
Vanessa Chicarino, C{\'e}lio Albuquerque, Emanuel Jesus, and Antonio Rocha.
\newblock On the detection of selfish mining and stalker attacks in blockchain
  networks.
\newblock {\em Annals of Telecommunications}, 75:143--152, 2020.

\bibitem{otsuki2021impact}
Kai Otsuki, Ryuya Nakamura, and Kazuyuki Shudo.
\newblock Impact of saving attacks on blockchain consensus.
\newblock {\em IEEE Access}, 9:133011--133022, 2021.

\bibitem{bordel2021denial}
Borja Bordel, Ram{\'o}n Alcarria, and Tom{\'a}s Robles.
\newblock Denial of chain: Evaluation and prediction of a novel cyberattack in
  blockchain-supported systems.
\newblock {\em Future Generation Computer Systems}, 116:426--439, 2021.

\bibitem{moubarak2018blockchain}
Joanna Moubarak, Eric Filiol, and Maroun Chamoun.
\newblock On blockchain security and relevant attacks.
\newblock In {\em 2018 IEEE Middle East and North Africa Communications
  Conference (MENACOMM)}, pages 1--6, Los Alamitos, CA, USA, 2018. IEEE.

\bibitem{stucke2022simulation}
Zachary Stucke, Theodoros Constantinides, and John Cartlidge.
\newblock Simulation of front-running attacks and privacy mitigations in
  ethereum blockchain.
\newblock In {\em 34th European Modeling and Simulation Symposium, EMSS 2022},
  page 041. Caltek, 2022.

\bibitem{wuthier2021proof}
Simeon Wuthier and Sang-Yoon Chang.
\newblock Proof-of-work network simulator for blockchain and cryptocurrency
  research.
\newblock In {\em 2021 IEEE 41st International Conference on Distributed
  Computing Systems (ICDCS)}, pages 1098--1101, Los Alamitos, CA, USA, 2021.
  IEEE.

\bibitem{khan2020simulation}
Kashif~Mehboob Khan, Junaid Arshad, and Muhammad~Mubashir Khan.
\newblock Simulation of transaction malleability attack for blockchain-based
  e-voting.
\newblock {\em Computers \& Electrical Engineering}, 83:106583, 2020.

\bibitem{kotenko2010agent}
Igor Kotenko, Alexey Konovalov, and Andrey Shorov.
\newblock Agent-based modeling and simulation of botnets and botnet defense.
\newblock In {\em Conference on Cyber Conflict. CCD COE Publications. Tallinn,
  Estonia}, pages 21--44. Citeseer, 2010.

\bibitem{gupta2002denial}
Vikram Gupta, Srikanth Krishnamurthy, and Michalis Faloutsos.
\newblock Denial of service attacks at the mac layer in wireless ad hoc
  networks.
\newblock In {\em MILCOM 2002. Proceedings}, volume~2, pages 1118--1123, Los
  Alamitos, CA, USA, 2002. IEEE.

\bibitem{kalluri2016simulation}
Rajesh Kalluri, Lagineni Mahendra, RK~Senthil Kumar, and GL~Ganga Prasad.
\newblock Simulation and impact analysis of denial-of-service attacks on power
  scada.
\newblock In {\em 2016 national power systems conference (NPSC)}, pages 1--5,
  Los Alamitos, CA, USA, 2016. IEEE.

\bibitem{sabri2021slowloris}
Shima Sabri, Noraini Ismail, and Amir Hazzim.
\newblock Slowloris dos attack based simulation.
\newblock In {\em IOP Conference series: materials science and engineering},
  volume 1062, page 012029. IOP Publishing, 2021.

\bibitem{oktian2014mitigating}
Yustus~Eko Oktian, SangGon Lee, and HoonJae Lee.
\newblock Mitigating denial of service (dos) attacks in openflow networks.
\newblock In {\em 2014 International Conference on Information and
  Communication Technology Convergence (ICTC)}, pages 325--330, Los Alamitos,
  CA, USA, 2014. IEEE.

\bibitem{bogdanoski2013analysis}
Mitko Bogdanoski, Tomislav Suminoski, and Aleksandar Risteski.
\newblock Analysis of the syn flood dos attack.
\newblock {\em International Journal of Computer Network and Information
  Security (IJCNIS)}, 5(8):1--11, 2013.

\bibitem{zhang2017opnet}
Jianmin Zhang, Yi~Chen, Naizheng Jin, Lianquan Hou, and Qianzhi Zhang.
\newblock Opnet based simulation modeling and analysis of dos attack for
  digital substation.
\newblock In {\em 2017 IEEE Power \& Energy Society General Meeting}, pages
  1--5, Los Alamitos, CA, USA, 2017. IEEE.

\bibitem{lian2007simulation}
Zhaotong Lian and Jian Lin.
\newblock Simulation analysis of the dos attack in internet service.
\newblock In {\em 2007 International Conference on Wireless Communications,
  Networking and Mobile Computing}, pages 6298--6301, Los Alamitos, CA, USA,
  2007. IEEE.

\bibitem{asri2015impact}
Satin Asri and Bernardi Pranggono.
\newblock Impact of distributed denial-of-service attack on advanced metering
  infrastructure.
\newblock {\em Wireless Personal Communications}, 83:2211--2223, 2015.

\bibitem{ni2018cyber}
Ming Ni, Yusheng Xue, Heqin Tong, and Manli Li.
\newblock A cyber physical power system co-simulation platform.
\newblock In {\em 2018 Workshop on Modeling and Simulation of Cyber-Physical
  Energy Systems (MSCPES)}, pages 1--5, Los Alamitos, CA, USA, 2018. IEEE.

\bibitem{nagarjun2013simulation}
PMD Nagarjun, V~Anil Kumar, Ch~Aswani Kumar, and Ahkshaey Ravi.
\newblock Simulation and analysis of rts/cts dos attack variants in 802.11
  networks.
\newblock In {\em 2013 International Conference on Pattern Recognition,
  Informatics and Mobile Engineering}, pages 258--263, Los Alamitos, CA, USA,
  2013. IEEE.

\bibitem{aslam2008802}
Baber Aslam, Monis Akhlaq, and Shoad~A Khan.
\newblock 802.11 disassociation dos attack simulation using verilog.
\newblock {\em WSEAS Transactions on Communications}, 7(3):198--206, 2008.

\bibitem{razak2002network}
Shabana Razak, Mian Zhou, and Sheau-Dong Lang.
\newblock Network intrusion simulation using opnet.
\newblock In {\em Proceedings of 2002 OPNETWORKS Conference}. Citeseer, 2002.

\bibitem{shin2023beyond}
Jeongkeun Shin, L~Richard Carley, Geoffrey~B Dobson, and Kathleen~M Carley.
\newblock Beyond accuracy: Cybersecurity resilience evaluation of intrusion
  detection system against dos attacks using agent-based simulation.
\newblock In {\em 2023 Winter Simulation Conference (WSC)}, pages 118--129, Los
  Alamitos, CA, USA, 2023. IEEE.

\bibitem{lotfy2013performance}
Poussy~A Lotfy and Marianne~A Azer.
\newblock Performance evaluation of aodv under dos attacks.
\newblock In {\em 6th Joint IFIP Wireless and Mobile Networking Conference
  (WMNC)}, pages 1--4, Los Alamitos, CA, USA, 2013. IEEE.

\bibitem{furfaro2015simulation}
Angelo Furfaro, Giovanna Malena, Lorena Molina, and Andrea Parise.
\newblock A simulation model for the analysis of ddos amplification attacks.
\newblock In {\em 2015 17th UKSim-AMSS International Conference on Modelling
  and Simulation (UKSim)}, pages 267--272, Los Alamitos, CA, USA, 2015. IEEE.

\bibitem{yan2011cyber}
Jie Yan, Chen-Ching Liu, and Manimaran Govindarasu.
\newblock Cyber intrusion of wind farm scada system and its impact analysis.
\newblock In {\em 2011 IEEE/PES Power Systems Conference and Exposition}, pages
  1--6, 2011.

\bibitem{choi2020multi}
In-Sun Choi, Junho Hong, and Tae-Wan Kim.
\newblock Multi-agent based cyber attack detection and mitigation for
  distribution automation system.
\newblock {\em IEEE Access}, 8:183495--183504, 2020.

\bibitem{fritz2019simulation}
Jared~J Fritz, Joseph Sagisi, John James, Aaron~St Leger, Kyle King, and Kate~J
  Duncan.
\newblock Simulation of man in the middle attack on smart grid testbed.
\newblock In {\em 2019 SoutheastCon}, pages 1--6, Los Alamitos, CA, USA, 2019.
  IEEE.

\bibitem{abdo2024vehicular}
Ahmed Abdo, Guoyuan Wu, and Nael Abu-Ghazaleh.
\newblock Vehicular secure network open simulator (vesnos): A cyber-security
  oriented co-simulation platform for connected and automated driving.
\newblock In {\em Proceedings of the 38th ACM SIGSIM Conference on Principles
  of Advanced Discrete Simulation}, SIGSIM-PADS '24, page 108–118, New York,
  NY, USA, 2024. Association for Computing Machinery.

\bibitem{patil2023review}
Sonali Patil, Mandaar Rao, Lavitra Misal, Darshan Phaldesai, and Kishor
  Shivsharan.
\newblock A review of the ow asp top 10 web application security risks and best
  practices for mitigating these risks.
\newblock In {\em 2023 7th International Conference On Computing,
  Communication, Control And Automation (ICCUBEA)}, pages 1--8, Los Alamitos,
  CA, USA, 2023. IEEE.

\bibitem{folan2023cybersecurity}
Sean Folan and Yunsheng Wang.
\newblock Cybersecurity simulator for connected and autonomous vehicles.
\newblock In {\em Proceedings of the Twenty-Fourth International Symposium on
  Theory, Algorithmic Foundations, and Protocol Design for Mobile Networks and
  Mobile Computing}, MobiHoc '23, page 430–435, New York, NY, USA, 2023.
  Association for Computing Machinery.

\bibitem{lingaraju2021simulation}
Kaushik Lingaraju, Jianzhong Gui, Brian~K Johnson, and Yacine Chakhchoukh.
\newblock Simulation of the effect of false data injection attacks on scada
  using pscad/emtdc.
\newblock In {\em 2020 52nd North American Power Symposium (NAPS)}, pages 1--5,
  Los Alamitos, CA, USA, 2021. IEEE.

\bibitem{kushal2018risk}
Tazim Ridwan~Billah Kushal, Kexing Lai, and Mahesh~S Illindala.
\newblock Risk-based mitigation of load curtailment cyber attack using
  intelligent agents in a shipboard power system.
\newblock {\em IEEE Transactions on Smart Grid}, 10(5):4741--4750, 2018.

\bibitem{tobin2015simulating}
Patrick Tobin, A~Mahrouqi, Sameh Abdalla, and Tahar Kechadi.
\newblock Simulating sql-injection cyber-attacks using gns3.
\newblock {\em International Journal of Computer Theory and Engineering},
  8(3):213--217, 2015.

\bibitem{blancaflor2024strengthening}
Eric Blancaflor, Christian~James Cabral, Kathlene~Joy Dejito, Justine~Paul
  Erni, Carlos~Nickolai Nicolas, and Ronaldo Bernardo.
\newblock Strengthening and securing online business transactions: A study on
  protecting against phishing attacks and sql code injection methods using
  sqlmap, clifty and secure coding practices.
\newblock In {\em Proceedings of the 2024 The 6th World Symposium on Software
  Engineering (WSSE)}, WSSE '24, page 7–14, New York, NY, USA, 2024.
  Association for Computing Machinery.

\bibitem{leszczyna2010simulating}
Rafa{\l} Leszczyna, Igor Nai~Fovino, and Marcelo Masera.
\newblock Simulating malware with malsim.
\newblock {\em Journal in Computer Virology}, 6:65--75, 2010.

\bibitem{hosseini2016agent}
Soodeh Hosseini, Mohammad Abdollahi~Azgomi, and Adel Rahmani~Torkaman.
\newblock Agent-based simulation of the dynamics of malware propagation in
  scale-free networks.
\newblock {\em Simulation}, 92(7):709--722, 2016.

\bibitem{benomar2022agent}
Ziyad Benomar, Chaima Ghribi, Elie Cali, Alexander Hinsen, and Benedikt Jahnel.
\newblock Agent-based modeling and simulation for malware spreading in d2d
  networks.
\newblock In {\em Proceedings of the 21st International Conference on
  Autonomous Agents and Multiagent Systems}, AAMAS '22, page 91–99, Richland,
  SC, 2022. International Foundation for Autonomous Agents and Multiagent
  Systems.

\bibitem{hara2024extending}
Jun'ichiro Hara, Koichi Mouri, and Eiji Takimoto.
\newblock Extending network simulator ns-3 for analyzing iot malware.
\newblock In {\em Proceedings of the 2024 13th International Conference on
  Networks, Communication and Computing}, ICNCC '24, page 73–80, New York,
  NY, USA, 2025. Association for Computing Machinery.

\bibitem{tanaka2017modeling}
Hiroaki Tanaka and Shingo Yamaguchi.
\newblock On modeling and simulation of the behavior of iot malwares mirai and
  hajime.
\newblock In {\em 2017 IEEE International Symposium on Consumer Electronics
  (ISCE)}, pages 56--60, Los Alamitos, CA, USA, 2017. IEEE.

\bibitem{bhasin2013physical}
Shivam Bhasin, Jean-Luc Danger, Tarik Graba, Yves Mathieu, Daisuke Fujimoto,
  and Makoto Nagata.
\newblock Physical security evaluation at an early design-phase: A side-channel
  aware simulation methodology.
\newblock In {\em Proceedings of International Workshop on Engineering
  Simulations for Cyber-Physical Systems}, ES4CPS '14, page 13–20, New York,
  NY, USA, 2013. Association for Computing Machinery.

\bibitem{kumar2017efficient}
Amit Kumar, Cody Scarborough, Ali Yilmaz, and Michael Orshansky.
\newblock Efficient simulation of em side-channel attack resilience.
\newblock In {\em 2017 IEEE/ACM International Conference on Computer-Aided
  Design (ICCAD)}, pages 123--130, Los Alamitos, CA, USA, 2017. IEEE.

\bibitem{menichelli2008high}
Francesco Menichelli, Renato Menicocci, Mauro Olivieri, and Alessandro
  Trifiletti.
\newblock High-level side-channel attack modeling and simulation for
  security-critical systems on chips.
\newblock {\em IEEE Transactions on Dependable and Secure Computing},
  5(3):164--176, 2008.

\bibitem{koutiva2021agent}
I~Koutiva, G~Moraitis, and C~Makropoulos.
\newblock An agent-based modelling approach to assess risk in cyber-physical
  systems (cps).
\newblock In {\em Proceedings of the 17th International Conference on
  Environmental Science and Technology, Athens, Greece}, pages 1--4, Los
  Alamitos, CA, USA, 2021.

\bibitem{furfaro2017using}
Angelo Furfaro, Luciano Argento, Andrea Parise, and Antonio Piccolo.
\newblock Using virtual environments for the assessment of cybersecurity issues
  in iot scenarios.
\newblock {\em Simulation Modelling Practice and Theory}, 73:43--54, 2017.

\bibitem{kotenko2005agent}
IV~Kotenko and AV~Ulanov.
\newblock Agent-based simulation of ddos attacks and defense mechanisms.
\newblock {\em Journal of Computing}, 4(2):16--37, 2005.

\bibitem{poisel2013game}
Rainer Poisel, Marlies Rybnicek, and Simon Tjoa.
\newblock Game-based simulation of distributed denial of service (ddos) attack
  and defense mechanisms of critical infrastructures.
\newblock In {\em 2013 IEEE 27th International Conference on Advanced
  Information Networking and Applications (AINA)}, pages 114--120, Los
  Alamitos, CA, USA, 2013. IEEE.

\bibitem{renaud2013simpass}
Karen Renaud and Lewis Mackenzie.
\newblock Simpass: Quantifying the impact of password behaviours and policy
  directives on an organisation’s systems.
\newblock {\em Journal of Artificial Societies and Social Simulation}, 16(2),
  2013.

\bibitem{rausch2018modeling}
Michael Rausch, Ahmed Fawaz, Ken Keefe, and William~H. Sanders.
\newblock Modeling humans: A general agent model for the evaluation of
  security.
\newblock In Annabelle McIver and Andras Horvath, editors, {\em Quantitative
  Evaluation of Systems}, pages 373--388, Cham, 2018. Springer International
  Publishing.

\bibitem{chiong2008modelling}
Raymond Chiong and Sandeep Dhakal.
\newblock Modelling database security through agent-based simulation.
\newblock In {\em 2008 Second Asia International Conference on Modelling \&
  Simulation (AMS)}, pages 24--28, Los Alamitos, CA, USA, 2008. IEEE.

\bibitem{kotenko2003agent}
Igor Kotenko and Evgeny Mankov.
\newblock Agent-based modeling and simulation of computer network attacks.
\newblock In {\em Fourth International Workshop Agent-Based Simulation}, volume~4, 2003.

\bibitem{kannan2016modeling}
Uma Kannan, Rajendran Swamidurai, and David Umphress.
\newblock Modeling host osi layers cyber-attacks using system dynamics.
\newblock In {\em Proceedings of the 2016 International Conference on Security
  and Management (SAM)}, pages 96--100, Las Vegas, USA, 2016. CSREA Press.

\bibitem{genge2015system}
B{\'e}la Genge, Istv{\'a}n Kiss, and Piroska Haller.
\newblock A system dynamics approach for assessing the impact of cyber attacks
  on critical infrastructures.
\newblock {\em International Journal of Critical Infrastructure Protection},
  10:3--17, 2015.

\bibitem{medoh2022future}
Chuks Medoh and Arnesh Telukdarie.
\newblock The future of cybersecurity: A system dynamics approach.
\newblock {\em Procedia Computer Science}, 200:318--326, 2022.
\newblock 3rd International Conference on Industry 4.0 and Smart Manufacturing.

\bibitem{roscoe2020simulation}
Jonathan~Francis Roscoe, Oliver Baxandall, and Robert Hercock.
\newblock Simulation of malware propagation and effects in connected and
  autonomous vehicles.
\newblock In {\em 2020 International Conference on Computing, Electronics \&
  Communications Engineering (iCCECE)}, pages 57--62, Los Alamitos, CA, USA,
  2020. IEEE.

\bibitem{kharabsheh2024seir}
Mohammad Kharabsheh, Issa Al-aiash, Ala Mughaid, and Muder Almiani.
\newblock The seir model for predicting malware propagation in computer
  networks.
\newblock In {\em 2024 International Conference on Intelligent Computing,
  Communication, Networking and Services (ICCNS)}, pages 108--113, Los
  Alamitos, CA, USA, 2024. IEEE.

\bibitem{mayfield2019component}
Katia~P. Mayfield, Mikel~D. Petty, Tymaine~S. Whitaker, John~A. Bland, and
  Walter~A. Cantrell.
\newblock Component-based implementation of cyberattack simulation models.
\newblock In {\em Proceedings of the 2019 ACM Southeast Conference}, ACMSE '19,
  page 64–71, New York, NY, USA, 2019. Association for Computing Machinery.

\bibitem{tritilanunt2006using}
Suratose Tritilanunt, Colin Boyd, Ernest Foo, and Juan~Gonzalez Nieto.
\newblock Using coloured petri nets to simulate dos-resistant protocols.
\newblock In K~Jensen, editor, {\em 7th Workshop and Tutorial on Practical Use
  of Coloured Petri Nets and the CPN Tools}, pages 261--280, Denmark, 2006.
  University of Aarhus.

\bibitem{petty2022modeling}
Mikel~D. Petty, Tymaine~S. Whitaker, E.~Michael Bearss, John~A. Bland,
  Walter~Alan Cantrell, C.~Daniel Colvett, and Katia~P. Maxwell.
\newblock Modeling cyberattacks with extended petri nets.
\newblock In {\em Proceedings of the 2022 ACM Southeast Conference}, ACMSE '22,
  page 67–73, New York, NY, USA, 2022. Association for Computing Machinery.

\bibitem{lee2023simulation}
Dae-Hwi Lee, Chan-Min Kim, Hyun-Seok Song, Yong-Hee Lee, and Won-Sun Chung.
\newblock Simulation-based cybersecurity testing and evaluation method for
  connected car v2x application using virtual machine.
\newblock {\em Sensors}, 23(3):1421, 2023.

\bibitem{potteiger2017evaluating}
Bradley Potteiger, William Emfinger, Himanshu Neema, Xenofon Koutosukos,
  CheeYee Tang, and Keith Stouffer.
\newblock Evaluating the effects of cyber-attacks on cyber physical systems
  using a hardware-in-the-loop simulation testbed.
\newblock In {\em 2017 Resilience Week (RWS)}, pages 177--183, Los Alamitos,
  CA, USA, 2017. IEEE.

\bibitem{adhikari2016wams}
Uttam Adhikari, Thomas Morris, and Shengyi Pan.
\newblock Wams cyber-physical test bed for power system, cybersecurity study,
  and data mining.
\newblock {\em IEEE Transactions on Smart Grid}, 8(6):2744--2753, 2016.

\bibitem{hou2024simulation}
Yunhui Hou, Mengxiang Liu, and Na~Shen.
\newblock A simulation research method for data injection attacks on dc
  microgrids.
\newblock In {\em 2024 China Automation Congress (CAC)}, pages 2460--2465, Los
  Alamitos, CA, USA, 2024. IEEE.

\bibitem{sun2022hil}
Leigang Sun and Jin Zhang.
\newblock Hil-ddos: A ddos attack simulation system based on
  hardware-in-the-loop.
\newblock In {\em Proceedings of the 4th International Conference on Advanced
  Information Science and System}, AISS '22, New York, NY, USA, 2023.
  Association for Computing Machinery.

\bibitem{diaz2016simulation}
Alvaro Diaz and Pablo Sanchez.
\newblock Simulation of attacks for security in wireless sensor network.
\newblock {\em Sensors}, 16(11):1932, 2016.

\bibitem{rothbart2004high}
Klaus Rothbart, Ulrich Neffe, Ch~Steger, Reinhold Weiss, Edgar Rieger, and
  Andreas M{\"u}hlberger.
\newblock High level fault injection for attack simulation in smart cards.
\newblock In {\em 13th Asian Test Symposium}, pages 118--121, Los Alamitos, CA,
  USA, 2004. IEEE.

\bibitem{tatar2016impact}
Unal Tatar, Hayretdin Bahsi, and Adrian Gheorghe.
\newblock Impact assessment of cyber attacks: A quantification study on power
  generation systems.
\newblock In {\em 2016 11th System of Systems Engineering Conference (SoSE)},
  pages 1--6, Los Alamitos, CA, USA, 2016. IEEE.

\bibitem{meyur2020bayesian}
Rounak Meyur.
\newblock A bayesian attack tree based approach to assess cyber-physical
  security of power system.
\newblock In {\em 2020 IEEE Texas Power and Energy Conference (TPEC)}, pages
  1--6, Los Alamitos, CA, USA, 2020. IEEE.

\bibitem{futuransky2009}
Ariel Futoransky, Fernando Miranda, Jos\'{e} Orlicki, and Carlos Sarraute.
\newblock Simulating cyber-attacks for fun and profit.
\newblock In {\em Proceedings of the 2nd International Conference on Simulation
  Tools and Techniques}, Simutools '09, Brussels, BEL, 2009. ICST (Institute
  for Computer Sciences, Social-Informatics and Telecommunications
  Engineering).

\bibitem{liu2019modeling}
Wanping Liu.
\newblock Modeling ransomware spreading by a dynamic node-level method.
\newblock {\em {IEEE} Access}, 7:142224--142232, 2019.

\bibitem{sarigiannidis2017modeling}
Panagiotis Sarigiannidis, Eirini Karapistoli, and Anastasios~A Economides.
\newblock Modeling the internet of things under attack: A g-network approach.
\newblock {\em IEEE Internet of Things Journal}, 4(6):1964--1977, 2017.

\bibitem{erdHodi2021simulating}
L{\'a}szl{\'o} Erd{\H{o}}di, {\AA}vald~{\AA}slaugson Sommervoll, and
  Fabio~Massimo Zennaro.
\newblock Simulating sql injection vulnerability exploitation using q-learning
  reinforcement learning agents.
\newblock {\em Journal of Information Security and Applications}, 61:102903,
  2021.

\bibitem{murillo2020co}
Andres Murillo, Riccardo Taormina, Nils Tippenhauer, and Stefano Galelli.
\newblock Co-simulating physical processes and network data for high-fidelity
  cyber-security experiments.
\newblock In {\em Sixth Annual Industrial Control System Security (ICSS)
  Workshop}, ICSS 2020, page 13–20, New York, NY, USA, 2021. Association for
  Computing Machinery.

\bibitem{koutsoukos2017sure}
Xenofon Koutsoukos, Gabor Karsai, Aron Laszka, Himanshu Neema, Bradley
  Potteiger, Peter Volgyesi, Yevgeniy Vorobeychik, and Janos Sztipanovits.
\newblock Sure: A modeling and simulation integration platform for evaluation
  of secure and resilient cyber--physical systems.
\newblock {\em Proceedings of the IEEE}, 106(1):93--112, 2017.

\bibitem{cheng2022modeling}
Rongjun Cheng, Hao Lyu, Yaxing Zheng, and Hongxia Ge.
\newblock Modeling and stability analysis of cyberattack effects on
  heterogeneous intelligent traffic flow.
\newblock {\em Physica A: Statistical Mechanics and its Applications},
  604:127941, 2022.

\bibitem{fu2021modeling}
Yangyang Fu, Zheng O'Neill, Zhiyao Yang, Veronica Adetola, Jin Wen, Lingyu Ren,
  Tim Wagner, Qi~Zhu, and Terresa Wu.
\newblock Modeling and evaluation of cyber-attacks on grid-interactive
  efficient buildings.
\newblock {\em Applied Energy}, 303:117639, 2021.

\bibitem{ciancamerla2013modeling}
Ester Ciancamerla, Michele Minichino, and S~Palmieri.
\newblock Modeling cyber attacks on a critical infrastructure scenario.
\newblock In {\em IISA 2013}, pages 1--6, Los Alamitos, CA, USA, 2013. IEEE.

\bibitem{yao2024simulation}
Jianbo Yao.
\newblock Simulation platform architecture design for side channel attack.
\newblock In {\em 2024 IEEE 4th International Conference on Data Science and
  Computer Application (ICDSCA)}, pages 93--97, Los Alamitos, CA, USA, 2024.
  IEEE.

\bibitem{liu2018real}
Zengji Liu, Qi~Wang, Yi~Tang, and Ming Ni.
\newblock The real-time co-simulation platform with hardware-in-loop for
  cyber-attack in smart grid.
\newblock In {\em 2018 IEEE Innovative Smart Grid Technologies-Asia (ISGT
  Asia)}, pages 845--849, Los Alamitos, CA, USA, 2018. IEEE.

\bibitem{top2017simulation}
Philip Top, Eddy Banks, Peter~D Barnes, Seth Bromberger, Brian~M Kelley,
  Rafael~Rivera Soto, Benjamin Salazar, Steven~G Smith, Nathan Yee, and Mark
  Freund.
\newblock Simulation of a rtu cyber attack on a transformer bank.
\newblock In {\em 2017 IEEE Power \& Energy Society General Meeting}, pages
  1--5, Los Alamitos, CA, USA, 2017. IEEE.

\bibitem{de2020co}
Evandro de~Souza, Omid Ardakanian, and Ioanis Nikolaidis.
\newblock A co-simulation platform for evaluating cyber security and control
  applications in the smart grid.
\newblock In {\em ICC 2020 - 2020 IEEE International Conference on
  Communications (ICC)}, pages 1--7, Los Alamitos, CA, USA, 2020. IEEE.

\bibitem{vaughn2016addressing}
Rayford~B. Vaughn and Tommy Morris.
\newblock Addressing critical industrial control system cyber security concerns
  via high fidelity simulation.
\newblock In {\em Proceedings of the 11th Annual Cyber and Information Security
  Research Conference}, CISRC '16, New York, NY, USA, 2016. Association for
  Computing Machinery.

\bibitem{carraminana2023enhancing}
David Carramiñana, Ana~M. Bernardos, Juan~A. Besada, and José~R. Casar.
\newblock Enhancing healthcare infrastructure resilience through agent-based
  simulation methods.
\newblock {\em Computer Communications}, 234:108070, 2025.

\bibitem{serena2022security}
Luca Serena, Gabriele D’Angelo, and Stefano Ferretti.
\newblock Security analysis of distributed ledgers and blockchains through
  agent-based simulation.
\newblock {\em Simulation Modelling Practice and Theory}, 114:102413, 2022.

\bibitem{long2005denial}
Men Long, Chwan-Hwa Wu, and John~Y Hung.
\newblock Denial of service attacks on network-based control systems: impact
  and mitigation.
\newblock {\em IEEE Transactions on Industrial Informatics}, 1(2):85--96, 2005.

\bibitem{kouril2014cloud}
Daniel Kouril, Tom{\'a}{\v{s}} Rebok, Tomas Jirsik, Jakub Cegan, Martin Drasar,
  Martin Vizv{\'a}ry, and Jan Vykopal.
\newblock Cloud-based testbed for simulation of cyber attacks.
\newblock In {\em 2014 IEEE Network Operations and Management Symposium
  (NOMS)}, pages 1--6, Los Alamitos, CA, USA, 2014. IEEE.

\bibitem{bayir2020company}
Batuhan Bayir, Ibrahim~Berk Yalinkilic, Sebnem Bora, and Ozgu Can.
\newblock Company security assesment with agent based simulation.
\newblock In {\em 2020 Innovations in Intelligent Systems and Applications
  Conference (ASYU)}, pages 1--6, Los Alamitos, CA, USA, 2020. IEEE.

\bibitem{charitoudi2014agent}
Konstantinia Charitoudi and Andrew~JC Blyth.
\newblock An agent-based socio-technical approach to impact assessment for
  cyber defense.
\newblock {\em Information Security Journal: A Global Perspective},
  23(4-6):125--136, 2014.

\bibitem{garetto2003modeling}
Michele Garetto, Weibo Gong, and Don Towsley.
\newblock Modeling malware spreading dynamics.
\newblock In {\em IEEE INFOCOM 2003. Twenty-Second annual joint conference of
  the IEEE computer and communications societies (IEEE Cat. No. 03CH37428)},
  volume~3, pages 1869--1879, Los Alamitos, CA, USA, 2003. IEEE.

\bibitem{skeoch2022modelling}
Henry~RK Skeoch.
\newblock Modelling ransomware attacks using pomdps.
\newblock In {\em Workshop on the Economics of Information Security}, 2022.

\bibitem{roque2024implementation}
Maria Soledad~Carrasco Roque, Jean Paul~Garcia Chancahuana, and Roberto~Montero
  Flores.
\newblock Implementation of security controls for the treatment of malware
  using breach and attack simulation.
\newblock In {\em 2024 IEEE XXXI International Conference on Electronics,
  Electrical Engineering and Computing (INTERCON)}, pages 1--7, Los Alamitos,
  CA, USA, 2024. IEEE.

\bibitem{sanodiya2017attacks}
Rakesh~Kumar Sanodiya.
\newblock Dos attacks: A simulation study.
\newblock In {\em 2017 International Conference on Energy, Communication, Data
  Analytics and Soft Computing (ICECDS)}, pages 2553--2558, Los Alamitos, CA,
  USA, 2017. IEEE.

\bibitem{wagner2016towards}
Neal Wagner, Cem~{\c{S}} {\c{S}}ahin, Michael Winterrose, James Riordan, Jaime
  Pena, Diana Hanson, and William~W Streilein.
\newblock Towards automated cyber decision support: A case study on network
  segmentation for security.
\newblock In {\em 2016 IEEE Symposium Series on Computational Intelligence
  (SSCI)}, pages 1--10, Los Alamitos, CA, USA, 2016. IEEE.

\bibitem{albarakati2022multi}
J~Aiman Albarakati, Mohamed Azeroual, Younes Boujoudar, Lahcen EL~Iysaouy,
  Ayman Aljarbouh, Asifa Tassaddiq, and Hassane EL~Markhi.
\newblock Multi-agent-based fault location and cyber-attack detection in
  distribution system.
\newblock {\em Energies}, 16(1):224, 2022.

\bibitem{dutt2013cyber}
Varun Dutt, Young-Suk Ahn, and Cleotilde Gonzalez.
\newblock Cyber situation awareness: modeling detection of cyber attacks with
  instance-based learning theory.
\newblock {\em Human Factors}, 55(3):605--618, 2013.

\bibitem{fielder2016modelling}
Andrew Fielder, Tingting Li, and Chris Hankin.
\newblock Modelling cost-effectiveness of defenses in industrial control
  systems.
\newblock In Amund Skavhaug, J{\'e}r{\'e}mie Guiochet, and Friedemann Bitsch,
  editors, {\em Computer Safety, Reliability, and Security}, pages 187--200,
  Cham, 2016. Springer International Publishing.

\bibitem{grabis2019simulation}
J{\=a}nis Grabis and Arturs Rasnacis.
\newblock Simulation based evaluation and tuning of distributed fraud detection
  algorithm.
\newblock In {\em 2019 Winter Simulation Conference (WSC)}, pages 786--796, Los
  Alamitos, CA, USA, 2019. IEEE.

\bibitem{prasad2024simulation}
Suranga Prasad, Pramod Munaweera, Tharaka Hewa, Yushan Siriwardhana, and Mika
  Ylinattila.
\newblock Simulation of iiot-driven attack vectors on 5g core networks: Dataset
  generation and machine learning based detection.
\newblock In {\em Proceedings of the 14th International Conference on the
  Internet of Things}, IoT '24, page 184–187, New York, NY, USA, 2025.
  Association for Computing Machinery.

\bibitem{aljamal2024optimizing}
Mahmoud AlJamal, Ala Mughaid, Hani Bani-Salameh, Shadi Alzubi, Laith Abualigah,
  et~al.
\newblock Optimizing risk mitigation: A simulation-based model for detecting
  fake iot clients in smart city environments.
\newblock {\em Sustainable Computing: Informatics and Systems}, 43:101019,
  2024.

\bibitem{de2024ddoshield}
Simona De~Vivo, Islam Obaidat, Dong Dai, and Pietro Liguori.
\newblock Ddoshield-iot: A testbed for simulating and lightweight detection of
  iot botnet ddos attacks.
\newblock In {\em 2024 54th Annual IEEE/IFIP International Conference on
  Dependable Systems and Networks Workshops (DSN-W)}, pages 1--8, Los Alamitos,
  CA, USA, 2024. IEEE.

\bibitem{jia2020flowguard}
Yizhen Jia, Fangtian Zhong, Arwa Alrawais, Bei Gong, and Xiuzhen Cheng.
\newblock Flowguard: An intelligent edge defense mechanism against iot ddos
  attacks.
\newblock {\em IEEE Internet of Things Journal}, 7(10):9552--9562, 2020.

\bibitem{wrana2022od1nf1st}
Michael~Maximilian Wrana, Marwa Elsayed, Karim Lounis, Ziad Mansour, Steven
  Ding, and Mohammad Zulkernine.
\newblock Od1nf1st: true skip intrusion detection and avionics network
  cyber-attack simulation.
\newblock {\em ACM Transactions on Cyber-Physical Systems (TCPS)}, 6(4):1--27,
  2022.

\bibitem{rimoli2024synthetic}
Gennaro~Pio Rimoli, Francesco Palmieri, and Massimo Ficco.
\newblock Synthetic threat dataset generation by uav fleet simulation.
\newblock In {\em The 14th International Defense and Homeland Security
  Simulation Workshop}, pages 1--8, 2024.

\bibitem{carrasco2024cybershield}
José Álvaro~Fernández Carrasco, Iñigo~Amonarriz Pagola, Raúl~Orduna
  Urrutia, and Rodrigo Román.
\newblock Cybershield: A competitive simulation environment for training ai in
  cybersecurity.
\newblock In {\em 2024 11th International Conference on Internet of Things:
  Systems, Management and Security (IOTSMS)}, pages 11--18, Los Alamitos, CA,
  USA, 2024. IEEE.

\bibitem{ashiku2020agent}
Lirim Ashiku and Cihan Dagli.
\newblock Agent based cybersecurity model for business entity risk assessment.
\newblock In {\em 2020 IEEE International Symposium on Systems Engineering
  (ISSE)}, pages 1--6, Los Alamitos, CA, USA, 2020. IEEE.

\bibitem{ali2024development}
Md~Hazrat~Ali, Asad Waqar~Malik, Nursultan Jyeniskhan, Muhammad Arif~Mahmood,
  Essam Shehab, and Frank Liou.
\newblock Development of digital twin for fdm printer with preventive
  cyber-attack and control algorithms.
\newblock {\em {IEEE} Access}, 12:193594--193606, 2024.

\bibitem{malekzadeh2011validating}
Mina Malekzadeh, Abdul Azim~Abdul Ghani, Shamala Subramaniam, and Jalil~M Desa.
\newblock Validating reliability of omnet++ in wireless networks dos attacks:
  Simulation vs. testbed.
\newblock {\em Int. J. Netw. Secur.}, 13(1):13--21, 2011.

\bibitem{eastman2017simulation}
David Eastman and Sathish~AP Kumar.
\newblock A simulation study to detect attacks on internet of things.
\newblock In {\em 2017 IEEE 15th Intl Conf on Dependable, Autonomic and Secure
  Computing, 15th Intl Conf on Pervasive Intelligence and Computing, 3rd Intl
  Conf on Big Data Intelligence and Computing and Cyber Science and Technology
  Congress (DASC/PiCom/DataCom/CyberSciTech)}, pages 645--650, Los Alamitos,
  CA, USA, 2017. IEEE.

\bibitem{hussain2003framework}
Alefiya Hussain, John Heidemann, and Christos Papadopoulos.
\newblock A framework for classifying denial of service attacks.
\newblock In {\em Proceedings of the 2003 Conference on Applications,
  Technologies, Architectures, and Protocols for Computer Communications},
  SIGCOMM '03, page 99–110, New York, NY, USA, 2003. Association for
  Computing Machinery.

\bibitem{zhang2021network}
Xiaoning Zhang, Hengwei Zhang, Chenwei Li, Pengyu Sun, Zhilin Liu, and Jindong
  Wang.
\newblock Network attack intention recognition based on signaling game model
  and netlogo simulation.
\newblock In {\em 2021 International Conference on Digital Society and
  Intelligent Systems (DSInS)}, pages 162--166, Los Alamitos, CA, USA, 2021.
  IEEE.

\bibitem{mashayekhi2024simulation}
Laila Mashayekhi and Michael~E Kuhl.
\newblock Simulation of low earth orbit satellite communication data for cyber
  attack detection.
\newblock In {\em 2024 Winter Simulation Conference (WSC)}, pages 2050--2057,
  Los Alamitos, CA, USA, 2024. IEEE.

\end{thebibliography}
\end{document}